\documentclass[9pt,twocolumn,floatfix,pra]{revtex4-1}
\usepackage{amsmath,amssymb,amsthm,mathrsfs,amsfonts,dsfont} 
\usepackage{amsmath,amssymb,amstext}
\usepackage{textcomp}
\usepackage{pbox}

\usepackage[export]{adjustbox}
\usepackage{tikz}
\usepackage{bm}
\usepackage{dcolumn}
\usepackage{booktabs}
\usepackage[scaled]{helvet}
\usepackage{sansmath}
\usepackage{graphicx}
\usepackage{transparent}
\usepackage{color}
\usepackage{gensymb}
\usepackage{multirow}
\usepackage{url}

\usepackage[colorlinks=true]{hyperref}
\usepackage[caption=false]{subfig}

\newcommand{\ket}[1]{\vert#1\rangle}
\newcommand{\bra}[1]{\langle#1\vert}

\newcommand{\nd}[0]{Nd$^{3+}$:Y$_2$SiO$_5$}

\newcommand{\suppmat}[0]{Appendix}

\newcommand{\ebits}[0]{$e$bits}

\newcommand{\Eofvalue}[0]{1.18(4)}

\begin{document}

\newcommand{\TitleName}{Quantification of multi-dimensional entanglement stored in a crystal}
\title{\TitleName}

\author{Alexey Tiranov$^{1}$}
\author{S\'ebastien Designolle$^{1}$}
\author{Emmanuel Zambrini Cruzeiro$^{1}$}
\author{Jonathan Lavoie$^{1,\dag}$}
\author{Nicolas Brunner$^{1}$}
\author{Mikael Afzelius$^{1}$}
\author{Marcus Huber$^{1,2}$}
\author{Nicolas Gisin$^{1}$}

\affiliation{$^{1}$Groupe de Physique Appliqu\'ee, Universit\'e de Gen\`eve, CH-1211 Gen\`eve, Switzerland}
\affiliation{$^{2}$Institute for Quantum Optics and Quantum Information, Austrian Academy of Sciences, A-1090 Vienna, Austria}

\begin{abstract}

The use of multidimensional entanglement opens new perspectives for quantum information processing. However, an important challenge in practice is to certify and characterize multidimensional entanglement from measurement data that is typically limited.
Here we report the certification and quantification of two-photon multi-dimensional energy-time entanglement between many temporal modes, after one photon has been stored in a crystal. We develop a method for entanglement quantification which makes use of only sparse data obtained with limited resources. This allows us to efficiently certify entanglement of formation of 1.18~\textit{e}bits after performing quantum storage. The theoretical methods we develop can be readily extended to a wide range of experimental platforms, while our experimental results demonstrate the suitability of energy-time multi-dimensional entanglement for a quantum repeater architecture.

\end{abstract}

\maketitle 


Quantum entanglement represents a key resource for quantum information processing, e.g. in quantum communications. Of particular interest is the possibility of using multi-dimensional entangled states, which are proven to outperform standard two-qubit entangled states for a wide range of applications. In particular, high-dimensional entanglement can increase the quantum communication  channel capacity~\cite{Bennett1999}, as well as enhance key rate and resilience to errors in quantum key distribution~\cite{Bechmann2000,Cerf2002,Sheridan2010}. Moreover, it is also relevant for the implementation of device-independent quantum communication protocols~\cite{Acin2007}, allowing for more robust Bell tests~\cite{Vertesi2010} and enhanced security~\cite{Huber2013a}.

In recent years a strong effort has been devoted to the experimental implementation of multi-dimensional entangled systems, in particular in the context of photonic experiments. Different degrees of freedom were considered, such as orbital angular momentum~\cite{Mair2001,Dada2011,Krenn2014}, frequency~\cite{Olislager2011,Bernhard2013,Xing2014,Jin2016}, spatial modes~\cite{Edgar2012,Fickler2014,Schaeff2015}, time-bins~\cite{DeRiedmatten2002,Stucki2005,Ikuta2016} and energy-time~\cite{Thew2004,Richart2012}. Several experiments also demonstrated the potential of multi-dimensional entanglement for quantum cryptography~\cite{Groblacher2006,AliKhan2007,Mirhosseini2015,Zhong2015}. For this time-bins and energy-time entangled systems are suitable for implementations using optical fibres~\cite{Marcikic2002}. 

While these works open promising perspectives, the use of multi-dimensional entanglement for practical and efficient quantum communications still faces important challenges. Unavoidable losses in optical fibers require the use of quantum repeater schemes featuring quantum memories in order to reach long distances~\cite{Sangouard2011}. First steps were taken in realizing quantum memories beyond qubits. Notable experiments demonstrated the storage of three-dimensional entanglement of orbital angular momentum~\cite{Zhou2015,Ding2016}, as well as the implementation of a temporal multimode quantum memory capable of storing multiple entangled two-qubit pairs~\cite{Tiranov2016a}, a key step for achieving efficient entanglement distribution~\cite{Simon2007}. 

Another important challenge consists of certifying and characterizing multi-dimensional entanglement. Indeed, the complexity of these systems (i.e. in terms of the number of parameters for characterizing their quantum state) renders usual methods, such as quantum state tomography, completely unpractical. More efficient techniques have been developed, based e.g. on compressed sensing~\cite{Gross2010,Tonolini2014}, but usually require partial prior knowledge of the state. In general the problem of developing reliable and efficient methods for characterizing high-dimensional entanglement based on experimentally accessible data, which is typically limited, is an active area of research~\cite{Giovannini2013,Howland2016,Erker2017}.

In the present work we address these challenges by demonstrating the characterization of multi-dimensional energy-time entanglement stored in a rare-earth ion-doped crystal based on very sparse data. We first develop a method for quantifying multi-dimensional entanglement based on the knowledge of the diagonal elements of the density matrix, and a few off-diagonal elements. In our experiment, this corresponds to measuring in the time-of-arrival basis, and observing the coherence between neighboring temporal modes. Based on the fact that the density matrix must be positive---in order to correspond to a valid quantum state---our method imposes strong constraints on other (not directly measurable) elements of the density matrix. Therefore, we can prove rigorous lower bounds on the entanglement of formation of the state, even though the available data is limited.

We demonstrate the practical relevance of the method in our quantum storage experiment, involving energy-time entanglement of a photon pair containing up to 9 temporal modes. In particular we certify that the quantum state after storage has an entanglement of formation of at least 1.18~\ebits{}. To the best of our knowledge, this is the highest value certified so far in any experiment (even without storage). These results demonstrate the potential of energy-time entanglement combined with multimode quantum memories for creating and certifying multi-dimensional entanglement on long distances.

\begin{figure}[t]
\includegraphics[width=0.5\textwidth]{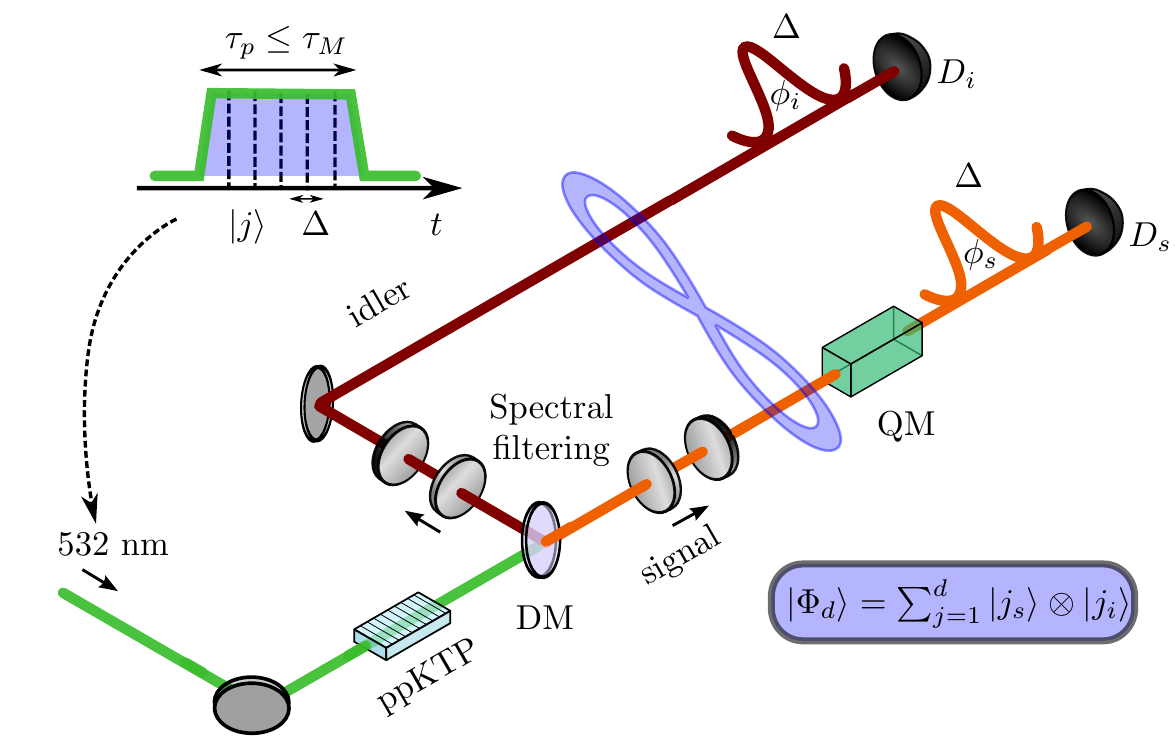}
\caption{Experimental setup. A pair of photons (signal and idler) is generated in a ppKTP waveguide via SPDC of a 532~nm pump photon. Both photons are spectrally filtered using optical cavities. Since the resulting coherence time of the photon pair is much smaller than the coherence time of the pump laser, this leads to the generation of two-photon energy-time entanglement. 
The signal photon is sent to a quantum memory (QM) based on a \nd{} crystal and stored for $\tau_M$=50~ns. The pump laser intensity is modulated using an acousto-optic modulator to generate a square pulse with duration $\tau_p$ smaller than the storage time of the~QM. 
Finally the photon pair is analyzed via an unbalanced interferometers, with controllable phases $\phi_s$ and $\phi_i$ and identical delays $\Delta= 5.5$~ns, and single-photon detectors ($D_s$ and $D_i$). Hence the entangled state generated and measured in our experiment can be compactly described by an entangled state of $d$ temporal modes of the form $\ket{\Phi_d}$. The experimental parameters allow for up to $d=9$ modes.
}
\label{fig:exp}
\end{figure}

We start by presenting our experimental scheme. Energy-time entanglement between two single photons at different wavelengths is generated using spontaneous parametric down conversion (SPDC). A monochromatic continuous-wave 532~nm laser pumps a nonlinear optical waveguide (periodically poled potassium titanyl phosphate (ppKTP) waveguide) to generate the signal and idler photons at 883~nm and 1338~nm, respectively (Fig.~\ref{fig:exp}). The two down-converted photons are created simultaneously and are well correlated in energy. However, the use of a monochromatic pump laser leads to an uncertainty on the photon pair creation time. This uncertainty is defined by the coherence time of the pump laser $ \sim 1$ ms and leads to energy-time entanglement between the two down-converted photons. 

The entangled photon pair is filtered down to 200~MHz which corresponds to a coherence time $\tau_c \approx 2.0$~ns (the details about the SPDC source can be found in~\cite{Clausen2014a}). The fact that $\tau_c$ is much smaller than the coherence time of the pump laser, combined with the detection scheme we use (see below), allow us to describe the entangled two-photon state as
\begin{align}
\ket{\Phi_d} = \frac{1}{\sqrt{d}}\sum_{j=1}^d \ket{j}_i \otimes \ket{j}_s\ ,
\label{eq:qudit}
\end{align}
where $\ket{j}_i$ ($\ket{j}_s$) denotes the state of the idler (signal) photon in temporal mode $j=1,...,d$. 

The signal photon is coupled to the quantum memory, which is based on a rare-earth ion-doped orthosilicate crystal, \nd{}, which is cooled down to 3~K. Photon storage is achieved via the atomic frequency comb (AFC) quantum memory protocol, implemented on the optical transition $^{4}I_{9/2}\longleftrightarrow^{4}F_{3/2}$ of Nd$^{3+}$ ions. The storage time of $\tau_M=50$~ns is predetermined, with an overall efficiency of 15\%. More details about the quantum memory can be found in~\cite{Clausen2011}. Here we use two-level AFC scheme which has a predetermined storage time and can be seen as temporal delay line.

Finally, local measurements are performed on each photon using unbalanced interferometers (see Fig.~\ref{fig:exp}). 
The delay $\Delta=5.5$~ns between the short and long arms of the interferometers is bigger than the coherence time of the photon pair $\tau_c$. In this case, the situation in which both photons passed through the short arm is indistinguishable from one where both photons travel through the long arm, leading to quantum interference in the coincidence rate~\cite{Franson1989}. In practice, two Michelson interferometers (bulk for the signal photon and fiber-based for the idler photon) with controllable phases ($\phi_s$ and $\phi_i$ on Fig.~\ref{fig:exp}) and identical delays $\Delta_s = \Delta_i = \Delta$ were implemented and actively phase stabilised~\cite{Tiranov2015a}. 

The experiment thus generates an energy-time entangled state between $d$ temporal modes, of the form \eqref{eq:qudit}, which can also be viewed as a post-selected time-bin entangled state. The maximum number of temporal modes that is possible to couple to the~QM is defined by its storage time~$\tau_M$. Thus the ratio~$\tau_M / \Delta$ corresponds to the maximum dimension of $d \sim 9$ for the state~\eqref{eq:qudit} which can be stored and certified in our experiment.

Our goal now is to characterize the multi-dimensional entanglement at the output of the  quantum memory by reconstructing part of the $d^2 \times d^2$ density matrix $ \rho$, with elements $\bra{j,k} \rho \ket{j',k'} = \text{Tr}[\rho (\ket{j}_i  \bra{j'}_i \otimes \ket{k}_s  \bra{k'}_s)]$. Note however, that the measurement information at our disposal is very limited, due to the simplicity of our measurement setup. Hence we can obtain only very few elements of $\rho$. Specifically, we can measure: i) the diagonal of the density matrix, i.e. terms $  \bra{j,k} \rho \ket{j,k}$, via the time-coincide measurement, and ii) the visibility $\mathcal{V}$ between two neighboring temporal modes, i.e. terms $  \bra{j,j} \rho \ket{j+1,j+1}$, via the interference measurements. Note that a full state reconstruction of $\rho$ would require the use of $d$ different interferometers, and is extremely cumbersome and unpractical. 

Nevertheless we will see that the limited information at our disposal is already enough to partly characterize the state, in particular leading to strong lower bounds on the entanglement of formation of $\rho$, $E_{oF}$. The latter is an operationally meaningful measure of entanglement, quantifying how much pure entanglement (counted in \ebits{}, i.e. the number of maximally entangled two-qubit pairs) is required in order to prepare $\rho$ via an arbitrary LOCC procedure. Following Ref.~\cite{Huber2013}, we have that 
\begin{equation}
E_{oF} \geq - \log_2(1-\frac{B^2}{2})\ ,
\label{eqn:eof}
\end{equation}
where we have defined the quantity 
\begin{equation}
B= \frac{2}{\sqrt{|C|}} \left(\sum_{(j,k)\in C \atop j < k}|\bra{j,j}\rho\ket{k,k}|-\sqrt{\bra{j,k}\rho\ket{j,k}\bra{k,j}\rho\ket{k,j}} \right ).
\label{eqn:b}
\end{equation}
Note that the indices $(j,k)$ are taken from a set $C$ that can be chosen at will. The quantity $B$ puts a lower bound on the concurrence of $\rho$~\cite{Wootters2001}. For a $d \times d$ maximally entangled pure state $\ket{\Phi_d}$ one has $B = \sqrt{2 (d-1)/d}$, leading to the tight bound $E_{oF} = \log_2(d)$.

\begin{figure}[t!]
\includegraphics[width=0.5\textwidth]{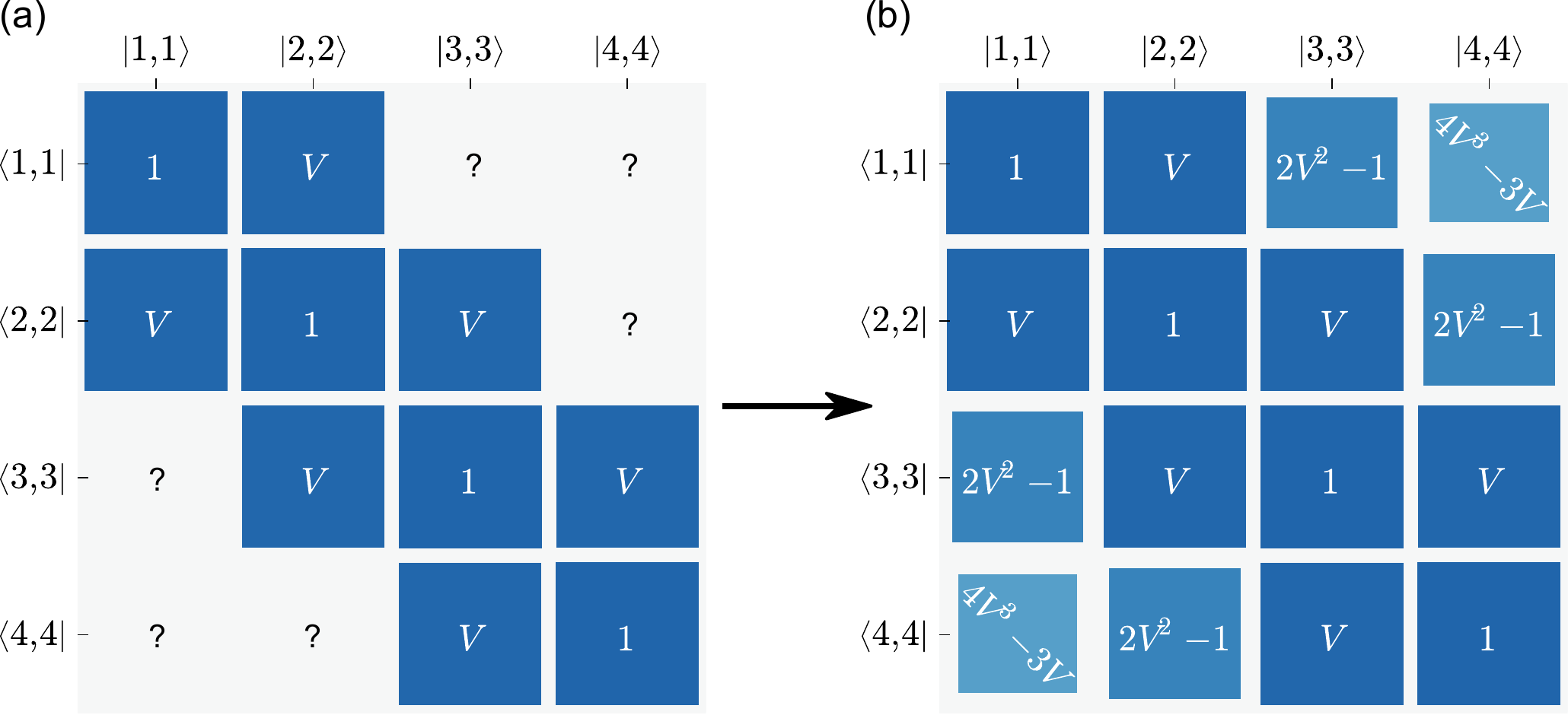}
\caption{Illustration of the method. Given a sub matrix where only the diagonal and first off diagonal are known (a), the method allows us to complete the matrix (b), giving lower bounds \eqref{eqn:bound_element} on all unknown elements based on positivity constraints. Finally this construction leads to a lower bound on the entanglement of formation via relation~(\ref{eqn:eof}).
}
\label{fig:dm}
\end{figure}

Note that the evaluation of $B$ requires only $O(d^2)$ elements of the density matrix, comparing to the total number of $d^4-1$. While the diagonal elements, i.e. $\bra{j,k}\rho\ket{j,k}$, can be estimated in the experiment (see below), measuring all coherence terms $\bra{j,j}\rho\ket{k,k}$ is still challenging and unpractical, as it requires many interferometers (with time delays $n \Delta$ with $n=2,...,d$) with controllable phases. 
Nevertheless we will see now that all unknown coherence terms (e.g. $\bra{j,j}\rho\ket{k,k}$ with $|k-j| \geq 2$) can in fact be efficiently lower bounded based only on accessible data. 

These bounds simply follow from the requirement of the density matrix $\rho$ to be positive semi-definite, i.e. representing a physical quantum state. We first notice that if a matrix is positive semi-definite, then it is also the case for its real part and all of its sub-matrices. Hence, the following sub-matrix of $\rho$ is positive semi-definite
\begin{equation}
\begin{pmatrix}
r_{1,1} & r_{1,2} & \cdots & r_{1,d} \\ 
r_{1,2} & r_{2,2} & \ddots & \vdots \\ 
\vdots & \ddots & \ddots & r_{d-1,d} \\ 
r_{1,d} & \cdots & r_{d-1,d} & r_{d,d}
\end{pmatrix}
\label{eqn:matrix}
\end{equation}
where $r_{j,k}=r_{k,j}=\Re(\bra{j,j}\rho\ket{k,k})$. From Sylvester's criterion it follows that every sub-determinant of a positive semi-definite matrix should be non-negative. In particular, the following determinant of any $3\times3$ sub-matrix of~(\ref{eqn:matrix}) is non-negative, i.e.
\begin{equation}
\begin{vmatrix}
r_{j,j} & r_{j,k} & r_{j,l} \\ 
r_{j,k} & r_{k,k} & r_{k,l} \\
r_{j,l} & r_{k,l} & r_{l,l}
\end{vmatrix} \geq 0
\ ,
\label{eqn:det3}
\end{equation}
for all $j<k<l$. We thus get the lower bound:
\begin{equation}
\small
r_{j,l} \geq   \frac{r_{j,k} r_{l,k}-\sqrt{(r_{j,j}r_{k,k}-r_{j,k}^2)(r_{k,k}r_{l,l}-r_{k,l}^2)}}{r_{k,k}}
\ .
\label{eqn:bound3}
\end{equation}
Notice that the square root in the above equation is real since its arguments are $2\times2$ sub-determinants of~(\ref{eqn:det3}) and therefore non-negative. Moreover, even if we do not know the exact value of $r_{j,k}$ or $r_{k,l}$, but only a non-negative lower bound on them, the formula~(\ref{eqn:bound3}) remains valid. This property allows us to iteratively compute a lower bound on every element of  the matrix \eqref{eqn:matrix}, based only on its diagonal and its first off-diagonal. 
Finally, we can lower bound $B$ and eventually the entanglement of formation $E_{oF}$ via inequality~(\ref{eqn:eof}). 
We emphasize that these lower bounds are general and rigorous, as they follow from the fact that the density matrix must be semi-definite positive, i.e. $\rho\geq 0$, in order to represent a valid quantum state.

\begin{figure*}[t!]
\centering
\includegraphics[width=0.67\textwidth]{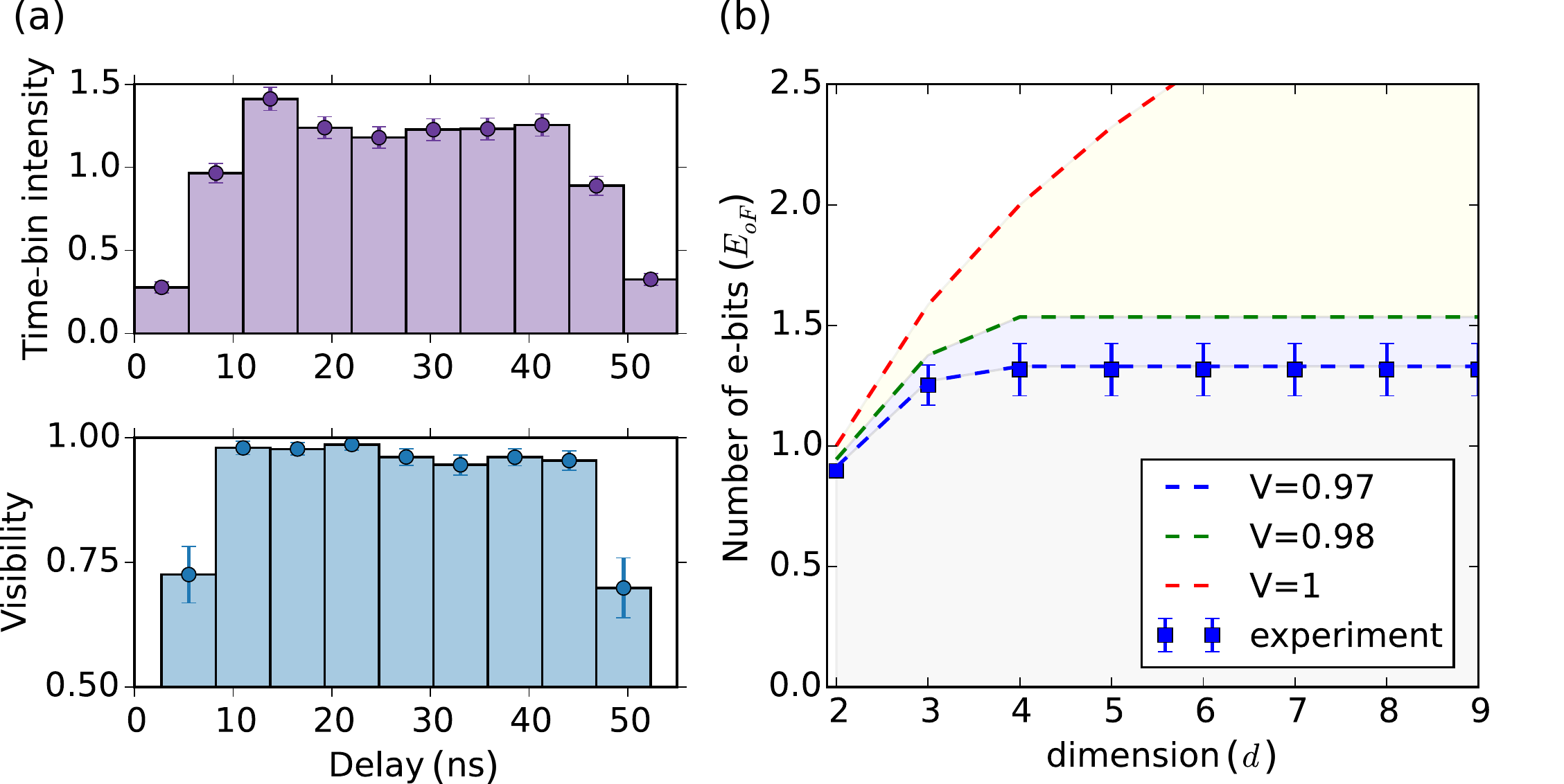}
\caption{Results for one experimental run. (a) The measured intensities in the time-of-arrival basis (diagonal elements $r_{j,j}$) and visibilities (first off-diagonal $r_{j,j+1}$) for $10$ temporal modes, separated by $\Delta = 5.5$~ns. (b) Lower bounds for the entanglement of formation (number of \ebits{}) as a function of the number $d$ of temporal modes taken into account when reconstructing the density matrix. Here the optimal value is $\sim$1.25(11) \ebits{}. The data shows good agreement with our model considering the measured visibility of $\mathcal{V}$ of 97\%. The case $\mathcal{V}=1$, corresponding to the maximally entangled state \eqref{eq:qudit}, gives $\log_2(d)$ \ebits{}. 
}
\label{fig:results}
\end{figure*}

Let us now focus on the situation of our experiment, for which we expect the following form of the density matrix (omitting normalization): $r_{j,j}=1$ for $j=1,...,d$ and $r_{j,j+1}= \mathcal{V}$ for $j=1,...,d-1$. The bounds on the first unknown off-diagonal elements read:
\begin{equation}
\begin{array}{rcl}
r_{j,j+2}  \geq  2\mathcal{V}^2-1  \quad, \quad r_{j,j+3}  \geq  \mathcal{V}(4\mathcal{V}^2-3) \, .
\end{array}
\label{eqn:bound_element}
\end{equation}
Hence the matrix \eqref{eqn:matrix}, containing initially many unknown elements, can be filled iteratively, as illustrated in Fig.~\ref{fig:dm}. Finally, by computing parameter $B$, we get a lower bound on the entanglement of formation depending on the visibility, see Fig.~\ref{fig:results}(b). In particular, for a perfect visibility $\mathcal{V}=1$, the only compatible state is the maximally entangled one \eqref{eq:qudit}, and the bound becomes tight, i.e. $E_{oF} = \log_2(d)$. See \suppmat{} B for more details.

Notice that the bounds become worse when one moves away from the diagonal. In fact, depending on the value of $\mathcal{V}$, the bound~(\ref{eqn:bound3}) becomes negative at some point, and thus the corresponding (and following) off-diagonal elements cannot be lower bounded anymore. Nevertheless, until that point, the bounds computed are useful. Notice also that we can play with the subset $C$ in Eq.~(\ref{eqn:b}) to improve the final bound on $E_{oF}$. This comes from the fact that, while taking a larger set $C$ makes the sum in Eq.~(\ref{eqn:b}) larger, the denominator $\sqrt{|C|}$ also grows. We find that in certain cases, better bounds on $E_{oF}$ are obtained when considering small sets $C$. 

To apply the above method to our experiment, we start by measuring the coherence between neighboring temporal modes, giving access to $r_{j,j+1}$. In order to do this, we use the two interferometers (Fig.~\ref{fig:exp}) to extract coherences between temporal modes $\ket{j}$ and $\ket{j+1}$. The phase of the idler interferometer $\phi_i$ is fixed while the phase of the signal interferometer $\phi_s$ is scanned over the interval $[0, 2\pi]$. For each time-bin the phase scan is done by measuring 15 points with 2 minutes per point. The coincidence rates are recorded, which correspond to local projections onto $e^{\phi_s+\phi_i}\ket{j,j}+\ket{j+1,j+1}$ for all $j=1,...,d-1$. The visibility values are extracted by comparing number of coincidences corresponding to constructive (maximum) and destructive (minimum) interference. Results are given in Fig.~\ref{fig:results}(a). Note that the visibilities for the first and the last temporal modes are lower due to a significant change of the intensity between the two neighboring modes. The average visibility for the central temporal modes is $\sim$97\%, and is limited by the interferometric stability and the multi-pair contribution from the SPDC process. 

We then measure correlations in the time basis, leading to the diagonal terms $r_{j,j}$ for $j=1,...,d$. For this, we block the short (or long) arm of the signal interferometer to project on states $\ket{j,j}$ (or $\ket{j+1,j+1}$) using a mechanical switch. The results for one of the measurements is depicted in Fig.~\ref{fig:results}(a).  The remaining terms of the diagonal of $\rho$, i.e. $\bra{j,k}\rho\ket{j,k}$ are also estimated. Essentially the only contributions to these elements are the multipair emission of the SPDC and noise of the detectors. Since these processes are independent of the temporal mode we assume that all diagonal terms $\bra{j,k}\rho\ket{j,k}$ are equal when $j\neq k$. Based on this assumption and using our interferometers we then measure contributions from neighboring modes $\bra{j,j+1}\rho\ket{j,j+1}$ which is approximately equal to $\approx 1\%$ and use these values for all other terms.

We analyze the data via the method discussed above in order to estimate the entanglement of formation of the state. We first lower bound each element in the submatrix \eqref{eqn:matrix}; details are given in the \suppmat{}. We consider all possible sub-matrices of $\rho$ (of different sizes) and keep the one leading to the best bound on $E_{oF}$, see Fig.~\ref{fig:results}(b). The maximum number of \ebits{} corresponds to the cases where both the measured visibilities and intensities are large and relatively constant. This is achieved by considering the central region of the pulse, excluding the edges where the intensity variation is limiting the visibility (Fig.~\ref{fig:results}(a)). 

Finally we obtain a lower bound for the entanglement of formation $E_{oF} \geq$ \Eofvalue{} \ebits{}, based on a dozen repetitions of the measurement procedure and analysis. The statistical error was measured for each experimental set and after propagated for all measurements. Moreover, this result also certifies a genuinely $3 \times 3$ entangled state, as any two-qubit state contains at most one $e$bit. More generally, our approach can be used to place lower bounds on the entanglement dimensionality given by $\log_2 (d) \geq E_{oF}$.

In the above analysis we certified a minimal degree of entanglement considering all possible quantum states (density matrices) compatible with our data. It is also relevant to estimate the entanglement based on a more physical model of our experiment. Indeed, this is expected to provide a much higher estimate of the entanglement, given that we consider here only quantum states of a specific form. Specifically, considering gaussian phase noise of the pump laser (with linewidth $\sim$1 kHz), we find that the visibility remains essentially constant for all temporal modes stored in the memory (see \suppmat{} for details). Hence we get that $r_{j,j+n} \approx \mathcal{V}$, where $\mathcal{V}$ is the measured visibility between two neighboring modes. This allows us to get a lower bound on the entanglement of formation of 2.6~\ebits{} from our measurement data which is limited by the visibility~$\mathcal{V}$.

In conclusion, we characterized multi-dimensional energy-time entanglement between two photons where one photon was stored in a crystal and the other photon is at telecom wavelength. In particular, we certified an entanglement of formation of \Eofvalue{} \ebits{}. For this we developed a general method for quantifying multidimensional entanglement. The method makes use of the fact the density matrix must be semi-definite positive, and provides strong lower bounds on the entanglement of formation, even when only sparse measurement data is available. The generality of our method may lead to applications in other physical platforms.
For instance, a recent follow-up of the present work certified high entanglement
in a purely photonic setup \cite{Martin2017}. Combined with the use of a quantum memory our approach offers promising perspectives for quantum communications based on multi-dimensional entanglement.

Due to its intrinsic temporal multimode capacity the AFC protocol fits well to realize a quantum repeater involving temporal multi-dimensional entanglement. For this a quantum memory that can retrieve photons on-demand using AFC spin-wave multimode storage~\cite{Jobez2016} or spectral multiplexing using multimode AFC delay lines \cite{Sinclair2014} could be used.

Finally, our method also serves as tool for certifying the dimensionality of entanglement. While we could certify $3 \times 3$ entanglement, higher dimensions could be reached by improving the state preparation and the measurement apparatus to achieve higher visibilities, or even use additional interferometers. Another interesting direction would be to perform device-independent tests of the degree of entanglement~\cite{Moroder2013} and its dimensionality~\cite{Brunner2008}. 

\section*{Acknowledgements}
We thank Florian Fr\"owis, F\'elix Bussi\`eres and Peter C. Strassmann for useful discussions.

\section*{Funding Information}
European Research Council (ERC-AG MEC). Swiss National Science Foundation (SNSF) (grant PP00P2-138917, Starting grant DIAQ, AMBIZIONE Z00P2-161351, and QSIT). Austrian Science Fund (FWF) through the START project Y879-N27. Natural Sciences and Engineering Research Council of Canada (NSERC).

\vspace{0.3cm}
{$^\dag$Present address: Department of Physics and Oregon Center for Optical Molecular \& Quantum Science, University of Oregon, Eugene, OR 97403, USA}
\vspace{0.3cm}




\bibliography{multidim_PRA.bbl}

\begin{thebibliography}{49}%
\makeatletter
\providecommand \@ifxundefined [1]{%
 \@ifx{#1\undefined}
}%
\providecommand \@ifnum [1]{%
 \ifnum #1\expandafter \@firstoftwo
 \else \expandafter \@secondoftwo
 \fi
}%
\providecommand \@ifx [1]{%
 \ifx #1\expandafter \@firstoftwo
 \else \expandafter \@secondoftwo
 \fi
}%
\providecommand \natexlab [1]{#1}%
\providecommand \enquote  [1]{``#1''}%
\providecommand \bibnamefont  [1]{#1}%
\providecommand \bibfnamefont [1]{#1}%
\providecommand \citenamefont [1]{#1}%
\providecommand \href@noop [0]{\@secondoftwo}%
\providecommand \href [0]{\begingroup \@sanitize@url \@href}%
\providecommand \@href[1]{\@@startlink{#1}\@@href}%
\providecommand \@@href[1]{\endgroup#1\@@endlink}%
\providecommand \@sanitize@url [0]{\catcode `\\12\catcode `\$12\catcode
  `\&12\catcode `\#12\catcode `\^12\catcode `\_12\catcode `\%12\relax}%
\providecommand \@@startlink[1]{}%
\providecommand \@@endlink[0]{}%
\providecommand \url  [0]{\begingroup\@sanitize@url \@url }%
\providecommand \@url [1]{\endgroup\@href {#1}{\urlprefix }}%
\providecommand \urlprefix  [0]{URL }%
\providecommand \Eprint [0]{\href }%
\providecommand \doibase [0]{http://dx.doi.org/}%
\providecommand \selectlanguage [0]{\@gobble}%
\providecommand \bibinfo  [0]{\@secondoftwo}%
\providecommand \bibfield  [0]{\@secondoftwo}%
\providecommand \translation [1]{[#1]}%
\providecommand \BibitemOpen [0]{}%
\providecommand \bibitemStop [0]{}%
\providecommand \bibitemNoStop [0]{.\EOS\space}%
\providecommand \EOS [0]{\spacefactor3000\relax}%
\providecommand \BibitemShut  [1]{\csname bibitem#1\endcsname}%
\let\auto@bib@innerbib\@empty
\bibitem [{\citenamefont {Bennett}\ \emph {et~al.}(1999)\citenamefont
  {Bennett}, \citenamefont {Shor}, \citenamefont {Smolin},\ and\ \citenamefont
  {Thapliyal}}]{Bennett1999}%
  \BibitemOpen
  \bibfield  {author} {\bibinfo {author} {\bibfnamefont {C.~H.}\ \bibnamefont
  {Bennett}}, \bibinfo {author} {\bibfnamefont {P.~W.}\ \bibnamefont {Shor}},
  \bibinfo {author} {\bibfnamefont {J.~A.}\ \bibnamefont {Smolin}}, \ and\
  \bibinfo {author} {\bibfnamefont {A.~V.}\ \bibnamefont {Thapliyal}},\ }\href
  {\doibase 10.1103/PhysRevLett.83.3081} {\bibfield  {journal} {\bibinfo
  {journal} {Phys. Rev. Lett.}\ }\textbf {\bibinfo {volume} {83}},\ \bibinfo
  {pages} {3081} (\bibinfo {year} {1999})}\BibitemShut {NoStop}%
\bibitem [{\citenamefont {Bechmann-Pasquinucci}\ and\ \citenamefont
  {Tittel}(2000)}]{Bechmann2000}%
  \BibitemOpen
  \bibfield  {author} {\bibinfo {author} {\bibfnamefont {H.}~\bibnamefont
  {Bechmann-Pasquinucci}}\ and\ \bibinfo {author} {\bibfnamefont
  {W.}~\bibnamefont {Tittel}},\ }\href {\doibase 10.1103/PhysRevA.61.062308}
  {\bibfield  {journal} {\bibinfo  {journal} {Phys. Rev. A}\ }\textbf {\bibinfo
  {volume} {61}},\ \bibinfo {pages} {062308} (\bibinfo {year}
  {2000})}\BibitemShut {NoStop}%
\bibitem [{\citenamefont {Cerf}\ \emph {et~al.}(2002)\citenamefont {Cerf},
  \citenamefont {Bourennane}, \citenamefont {Karlsson},\ and\ \citenamefont
  {Gisin}}]{Cerf2002}%
  \BibitemOpen
  \bibfield  {author} {\bibinfo {author} {\bibfnamefont {N.~J.}\ \bibnamefont
  {Cerf}}, \bibinfo {author} {\bibfnamefont {M.}~\bibnamefont {Bourennane}},
  \bibinfo {author} {\bibfnamefont {A.}~\bibnamefont {Karlsson}}, \ and\
  \bibinfo {author} {\bibfnamefont {N.}~\bibnamefont {Gisin}},\ }\href
  {\doibase 10.1103/PhysRevLett.88.127902} {\bibfield  {journal} {\bibinfo
  {journal} {Phys. Rev. Lett.}\ }\textbf {\bibinfo {volume} {88}},\ \bibinfo
  {pages} {127902} (\bibinfo {year} {2002})}\BibitemShut {NoStop}%
\bibitem [{\citenamefont {Sheridan}\ and\ \citenamefont
  {Scarani}(2010)}]{Sheridan2010}%
  \BibitemOpen
  \bibfield  {author} {\bibinfo {author} {\bibfnamefont {L.}~\bibnamefont
  {Sheridan}}\ and\ \bibinfo {author} {\bibfnamefont {V.}~\bibnamefont
  {Scarani}},\ }\href {\doibase 10.1103/PhysRevA.82.030301} {\bibfield
  {journal} {\bibinfo  {journal} {Phys. Rev. A}\ }\textbf {\bibinfo {volume}
  {82}},\ \bibinfo {pages} {030301} (\bibinfo {year} {2010})}\BibitemShut
  {NoStop}%
\bibitem [{\citenamefont {Ac\'{\i}n}\ \emph {et~al.}(2007)\citenamefont
  {Ac\'{\i}n}, \citenamefont {Brunner}, \citenamefont {Gisin}, \citenamefont
  {Massar}, \citenamefont {Pironio},\ and\ \citenamefont {Scarani}}]{Acin2007}%
  \BibitemOpen
  \bibfield  {author} {\bibinfo {author} {\bibfnamefont {A.}~\bibnamefont
  {Ac\'{\i}n}}, \bibinfo {author} {\bibfnamefont {N.}~\bibnamefont {Brunner}},
  \bibinfo {author} {\bibfnamefont {N.}~\bibnamefont {Gisin}}, \bibinfo
  {author} {\bibfnamefont {S.}~\bibnamefont {Massar}}, \bibinfo {author}
  {\bibfnamefont {S.}~\bibnamefont {Pironio}}, \ and\ \bibinfo {author}
  {\bibfnamefont {V.}~\bibnamefont {Scarani}},\ }\href {\doibase
  10.1103/PhysRevLett.98.230501} {\bibfield  {journal} {\bibinfo  {journal}
  {Phys. Rev. Lett.}\ }\textbf {\bibinfo {volume} {98}},\ \bibinfo {pages}
  {230501} (\bibinfo {year} {2007})}\BibitemShut {NoStop}%
\bibitem [{\citenamefont {V\'ertesi}\ \emph {et~al.}(2010)\citenamefont
  {V\'ertesi}, \citenamefont {Pironio},\ and\ \citenamefont
  {Brunner}}]{Vertesi2010}%
  \BibitemOpen
  \bibfield  {author} {\bibinfo {author} {\bibfnamefont {T.}~\bibnamefont
  {V\'ertesi}}, \bibinfo {author} {\bibfnamefont {S.}~\bibnamefont {Pironio}},
  \ and\ \bibinfo {author} {\bibfnamefont {N.}~\bibnamefont {Brunner}},\ }\href
  {\doibase 10.1103/PhysRevLett.104.060401} {\bibfield  {journal} {\bibinfo
  {journal} {Phys. Rev. Lett.}\ }\textbf {\bibinfo {volume} {104}},\ \bibinfo
  {pages} {060401} (\bibinfo {year} {2010})}\BibitemShut {NoStop}%
\bibitem [{\citenamefont {Huber}\ and\ \citenamefont
  {Paw\l{}owski}(2013)}]{Huber2013a}%
  \BibitemOpen
  \bibfield  {author} {\bibinfo {author} {\bibfnamefont {M.}~\bibnamefont
  {Huber}}\ and\ \bibinfo {author} {\bibfnamefont {M.}~\bibnamefont
  {Paw\l{}owski}},\ }\href {\doibase 10.1103/PhysRevA.88.032309} {\bibfield
  {journal} {\bibinfo  {journal} {Phys. Rev. A}\ }\textbf {\bibinfo {volume}
  {88}},\ \bibinfo {pages} {032309} (\bibinfo {year} {2013})}\BibitemShut
  {NoStop}%
\bibitem [{\citenamefont {Mair}\ \emph {et~al.}(2001)\citenamefont {Mair},
  \citenamefont {Vaziri}, \citenamefont {Weihs},\ and\ \citenamefont
  {Zeilinger}}]{Mair2001}%
  \BibitemOpen
  \bibfield  {author} {\bibinfo {author} {\bibfnamefont {A.}~\bibnamefont
  {Mair}}, \bibinfo {author} {\bibfnamefont {A.}~\bibnamefont {Vaziri}},
  \bibinfo {author} {\bibfnamefont {G.}~\bibnamefont {Weihs}}, \ and\ \bibinfo
  {author} {\bibfnamefont {A.}~\bibnamefont {Zeilinger}},\ }\href {\doibase
  10.1038/35085529} {\bibfield  {journal} {\bibinfo  {journal} {Nature}\
  }\textbf {\bibinfo {volume} {412}},\ \bibinfo {pages} {313} (\bibinfo {year}
  {2001})}\BibitemShut {NoStop}%
\bibitem [{\citenamefont {Dada}\ \emph {et~al.}(2011)\citenamefont {Dada},
  \citenamefont {Leach}, \citenamefont {Buller}, \citenamefont {Padgett},\ and\
  \citenamefont {Andersson}}]{Dada2011}%
  \BibitemOpen
  \bibfield  {author} {\bibinfo {author} {\bibfnamefont {A.}~\bibnamefont
  {Dada}}, \bibinfo {author} {\bibfnamefont {J.}~\bibnamefont {Leach}},
  \bibinfo {author} {\bibfnamefont {G.}~\bibnamefont {Buller}}, \bibinfo
  {author} {\bibfnamefont {M.}~\bibnamefont {Padgett}}, \ and\ \bibinfo
  {author} {\bibfnamefont {E.}~\bibnamefont {Andersson}},\ }\href {\doibase
  10.1038/nphys1996} {\bibfield  {journal} {\bibinfo  {journal} {Nature
  Physics}\ }\textbf {\bibinfo {volume} {7}},\ \bibinfo {pages} {677} (\bibinfo
  {year} {2011})}\BibitemShut {NoStop}%
\bibitem [{\citenamefont {Krenn}\ \emph {et~al.}(2014)\citenamefont {Krenn},
  \citenamefont {Huber}, \citenamefont {Fickler}, \citenamefont {Lapkiewicz},
  \citenamefont {Ramelow},\ and\ \citenamefont {Zeilinger}}]{Krenn2014}%
  \BibitemOpen
  \bibfield  {author} {\bibinfo {author} {\bibfnamefont {M.}~\bibnamefont
  {Krenn}}, \bibinfo {author} {\bibfnamefont {M.}~\bibnamefont {Huber}},
  \bibinfo {author} {\bibfnamefont {R.}~\bibnamefont {Fickler}}, \bibinfo
  {author} {\bibfnamefont {R.}~\bibnamefont {Lapkiewicz}}, \bibinfo {author}
  {\bibfnamefont {S.}~\bibnamefont {Ramelow}}, \ and\ \bibinfo {author}
  {\bibfnamefont {A.}~\bibnamefont {Zeilinger}},\ }\href {\doibase
  10.1073/pnas.1402365111} {\bibfield  {journal} {\bibinfo  {journal}
  {Proceedings of the National Academy of Sciences USA}\ }\textbf {\bibinfo
  {volume} {111}},\ \bibinfo {pages} {6243} (\bibinfo {year}
  {2014})}\BibitemShut {NoStop}%
\bibitem [{\citenamefont {Olislager}\ \emph {et~al.}(2010)\citenamefont
  {Olislager}, \citenamefont {Cussey}, \citenamefont {Nguyen}, \citenamefont
  {Emplit}, \citenamefont {Massar}, \citenamefont {Merolla},\ and\
  \citenamefont {Huy}}]{Olislager2011}%
  \BibitemOpen
  \bibfield  {author} {\bibinfo {author} {\bibfnamefont {L.}~\bibnamefont
  {Olislager}}, \bibinfo {author} {\bibfnamefont {J.}~\bibnamefont {Cussey}},
  \bibinfo {author} {\bibfnamefont {A.~T.}\ \bibnamefont {Nguyen}}, \bibinfo
  {author} {\bibfnamefont {P.}~\bibnamefont {Emplit}}, \bibinfo {author}
  {\bibfnamefont {S.}~\bibnamefont {Massar}}, \bibinfo {author} {\bibfnamefont
  {J.-M.}\ \bibnamefont {Merolla}}, \ and\ \bibinfo {author} {\bibfnamefont
  {K.~P.}\ \bibnamefont {Huy}},\ }\href {\doibase 10.1103/PhysRevA.82.013804}
  {\bibfield  {journal} {\bibinfo  {journal} {Phys. Rev. A}\ }\textbf {\bibinfo
  {volume} {82}},\ \bibinfo {pages} {013804} (\bibinfo {year}
  {2010})}\BibitemShut {NoStop}%
\bibitem [{\citenamefont {Bernhard}\ \emph {et~al.}(2013)\citenamefont
  {Bernhard}, \citenamefont {Bessire}, \citenamefont {Feurer},\ and\
  \citenamefont {Stefanov}}]{Bernhard2013}%
  \BibitemOpen
  \bibfield  {author} {\bibinfo {author} {\bibfnamefont {C.}~\bibnamefont
  {Bernhard}}, \bibinfo {author} {\bibfnamefont {B.}~\bibnamefont {Bessire}},
  \bibinfo {author} {\bibfnamefont {T.}~\bibnamefont {Feurer}}, \ and\ \bibinfo
  {author} {\bibfnamefont {A.}~\bibnamefont {Stefanov}},\ }\href {\doibase
  10.1103/PhysRevA.88.032322} {\bibfield  {journal} {\bibinfo  {journal} {Phys.
  Rev. A}\ }\textbf {\bibinfo {volume} {88}},\ \bibinfo {pages} {032322}
  (\bibinfo {year} {2013})}\BibitemShut {NoStop}%
\bibitem [{\citenamefont {Xing}\ \emph {et~al.}(2014)\citenamefont {Xing},
  \citenamefont {Feizpour}, \citenamefont {Hayat},\ and\ \citenamefont
  {Steinberg}}]{Xing2014}%
  \BibitemOpen
  \bibfield  {author} {\bibinfo {author} {\bibfnamefont {X.}~\bibnamefont
  {Xing}}, \bibinfo {author} {\bibfnamefont {A.}~\bibnamefont {Feizpour}},
  \bibinfo {author} {\bibfnamefont {A.}~\bibnamefont {Hayat}}, \ and\ \bibinfo
  {author} {\bibfnamefont {A.~M.}\ \bibnamefont {Steinberg}},\ }\href {\doibase
  10.1364/OE.22.025128} {\bibfield  {journal} {\bibinfo  {journal} {Optics
  express}\ }\textbf {\bibinfo {volume} {22}},\ \bibinfo {pages} {25128}
  (\bibinfo {year} {2014})}\BibitemShut {NoStop}%
\bibitem [{\citenamefont {Jin}\ \emph {et~al.}(2016)\citenamefont {Jin},
  \citenamefont {Shimizu}, \citenamefont {Fujiwara}, \citenamefont {Takeoka},
  \citenamefont {Wakabayashi}, \citenamefont {Yamashita}, \citenamefont {Miki},
  \citenamefont {Terai}, \citenamefont {Gerrits},\ and\ \citenamefont
  {Sasaki}}]{Jin2016}%
  \BibitemOpen
  \bibfield  {author} {\bibinfo {author} {\bibfnamefont {R.-B.}\ \bibnamefont
  {Jin}}, \bibinfo {author} {\bibfnamefont {R.}~\bibnamefont {Shimizu}},
  \bibinfo {author} {\bibfnamefont {M.}~\bibnamefont {Fujiwara}}, \bibinfo
  {author} {\bibfnamefont {M.}~\bibnamefont {Takeoka}}, \bibinfo {author}
  {\bibfnamefont {R.}~\bibnamefont {Wakabayashi}}, \bibinfo {author}
  {\bibfnamefont {T.}~\bibnamefont {Yamashita}}, \bibinfo {author}
  {\bibfnamefont {S.}~\bibnamefont {Miki}}, \bibinfo {author} {\bibfnamefont
  {H.}~\bibnamefont {Terai}}, \bibinfo {author} {\bibfnamefont
  {T.}~\bibnamefont {Gerrits}}, \ and\ \bibinfo {author} {\bibfnamefont
  {M.}~\bibnamefont {Sasaki}},\ }\href
  {http://stacks.iop.org/2058-9565/1/i=1/a=015004} {\bibfield  {journal}
  {\bibinfo  {journal} {Quantum Science and Technology}\ }\textbf {\bibinfo
  {volume} {1}},\ \bibinfo {pages} {015004} (\bibinfo {year}
  {2016})}\BibitemShut {NoStop}%
\bibitem [{\citenamefont {Edgar}\ \emph {et~al.}(2012)\citenamefont {Edgar},
  \citenamefont {Tasca}, \citenamefont {Izdebski}, \citenamefont {Warburton},
  \citenamefont {Leach}, \citenamefont {Agnew}, \citenamefont {Buller},
  \citenamefont {Boyd},\ and\ \citenamefont {Padgett}}]{Edgar2012}%
  \BibitemOpen
  \bibfield  {author} {\bibinfo {author} {\bibfnamefont {M.}~\bibnamefont
  {Edgar}}, \bibinfo {author} {\bibfnamefont {D.}~\bibnamefont {Tasca}},
  \bibinfo {author} {\bibfnamefont {F.}~\bibnamefont {Izdebski}}, \bibinfo
  {author} {\bibfnamefont {R.}~\bibnamefont {Warburton}}, \bibinfo {author}
  {\bibfnamefont {J.}~\bibnamefont {Leach}}, \bibinfo {author} {\bibfnamefont
  {M.}~\bibnamefont {Agnew}}, \bibinfo {author} {\bibfnamefont
  {G.}~\bibnamefont {Buller}}, \bibinfo {author} {\bibfnamefont
  {R.}~\bibnamefont {Boyd}}, \ and\ \bibinfo {author} {\bibfnamefont
  {M.}~\bibnamefont {Padgett}},\ }\href {\doibase 10.1038/ncomms1988}
  {\bibfield  {journal} {\bibinfo  {journal} {Nature Communications}\ }\textbf
  {\bibinfo {volume} {3}},\ \bibinfo {pages} {984} (\bibinfo {year}
  {2012})}\BibitemShut {NoStop}%
\bibitem [{\citenamefont {Fickler}\ \emph {et~al.}(2014)\citenamefont
  {Fickler}, \citenamefont {Lapkiewicz}, \citenamefont {Huber}, \citenamefont
  {Lavery}, \citenamefont {Padgett},\ and\ \citenamefont
  {Zeilinger}}]{Fickler2014}%
  \BibitemOpen
  \bibfield  {author} {\bibinfo {author} {\bibfnamefont {R.}~\bibnamefont
  {Fickler}}, \bibinfo {author} {\bibfnamefont {R.}~\bibnamefont {Lapkiewicz}},
  \bibinfo {author} {\bibfnamefont {M.}~\bibnamefont {Huber}}, \bibinfo
  {author} {\bibfnamefont {M.~P.}\ \bibnamefont {Lavery}}, \bibinfo {author}
  {\bibfnamefont {M.~J.}\ \bibnamefont {Padgett}}, \ and\ \bibinfo {author}
  {\bibfnamefont {A.}~\bibnamefont {Zeilinger}},\ }\href {\doibase
  10.1038/ncomms5502} {\bibfield  {journal} {\bibinfo  {journal} {Nature
  Communications}\ }\textbf {\bibinfo {volume} {5}},\ \bibinfo {pages} {5}
  (\bibinfo {year} {2014})}\BibitemShut {NoStop}%
\bibitem [{\citenamefont {Schaeff}\ \emph {et~al.}(2015)\citenamefont
  {Schaeff}, \citenamefont {Polster}, \citenamefont {Huber}, \citenamefont
  {Ramelow},\ and\ \citenamefont {Zeilinger}}]{Schaeff2015}%
  \BibitemOpen
  \bibfield  {author} {\bibinfo {author} {\bibfnamefont {C.}~\bibnamefont
  {Schaeff}}, \bibinfo {author} {\bibfnamefont {R.}~\bibnamefont {Polster}},
  \bibinfo {author} {\bibfnamefont {M.}~\bibnamefont {Huber}}, \bibinfo
  {author} {\bibfnamefont {S.}~\bibnamefont {Ramelow}}, \ and\ \bibinfo
  {author} {\bibfnamefont {A.}~\bibnamefont {Zeilinger}},\ }\href {\doibase
  10.1364/OPTICA.2.000523} {\bibfield  {journal} {\bibinfo  {journal} {Optica}\
  }\textbf {\bibinfo {volume} {2}},\ \bibinfo {pages} {7} (\bibinfo {year}
  {2015})}\BibitemShut {NoStop}%
\bibitem [{\citenamefont {de~Riedmatten}\ \emph {et~al.}(2002)\citenamefont
  {de~Riedmatten}, \citenamefont {Marcikic}, \citenamefont {Zbinden},\ and\
  \citenamefont {Gisin}}]{DeRiedmatten2002}%
  \BibitemOpen
  \bibfield  {author} {\bibinfo {author} {\bibfnamefont {H.}~\bibnamefont
  {de~Riedmatten}}, \bibinfo {author} {\bibfnamefont {I.}~\bibnamefont
  {Marcikic}}, \bibinfo {author} {\bibfnamefont {H.}~\bibnamefont {Zbinden}}, \
  and\ \bibinfo {author} {\bibfnamefont {N.}~\bibnamefont {Gisin}},\ }\href
  {http://arxiv.org/abs/quant-ph/0204165} {\bibfield  {journal} {\bibinfo
  {journal} {Quant. Inf. Comp.}\ }\textbf {\bibinfo {volume} {2}},\ \bibinfo
  {pages} {425} (\bibinfo {year} {2002})}\BibitemShut {NoStop}%
\bibitem [{\citenamefont {Stucki}\ \emph {et~al.}(2005)\citenamefont {Stucki},
  \citenamefont {Zbinden},\ and\ \citenamefont {Gisin}}]{Stucki2005}%
  \BibitemOpen
  \bibfield  {author} {\bibinfo {author} {\bibfnamefont {D.}~\bibnamefont
  {Stucki}}, \bibinfo {author} {\bibfnamefont {H.}~\bibnamefont {Zbinden}}, \
  and\ \bibinfo {author} {\bibfnamefont {N.}~\bibnamefont {Gisin}},\ }\href
  {\doibase 10.1080/09500340500283821} {\bibfield  {journal} {\bibinfo
  {journal} {Journal of Modern Optics}\ }\textbf {\bibinfo {volume} {52}},\
  \bibinfo {pages} {2637} (\bibinfo {year} {2005})}\BibitemShut {NoStop}%
\bibitem [{\citenamefont {Ikuta}\ and\ \citenamefont
  {Takesue}(2016)}]{Ikuta2016}%
  \BibitemOpen
  \bibfield  {author} {\bibinfo {author} {\bibfnamefont {T.}~\bibnamefont
  {Ikuta}}\ and\ \bibinfo {author} {\bibfnamefont {H.}~\bibnamefont
  {Takesue}},\ }\href {\doibase 10.1103/PhysRevA.93.022307} {\bibfield
  {journal} {\bibinfo  {journal} {Phys. Rev. A}\ }\textbf {\bibinfo {volume}
  {93}},\ \bibinfo {pages} {022307} (\bibinfo {year} {2016})}\BibitemShut
  {NoStop}%
\bibitem [{\citenamefont {Thew}\ \emph {et~al.}(2004)\citenamefont {Thew},
  \citenamefont {Ac{\'{i}}n}, \citenamefont {Zbinden},\ and\ \citenamefont
  {Gisin}}]{Thew2004}%
  \BibitemOpen
  \bibfield  {author} {\bibinfo {author} {\bibfnamefont {R.~T.}\ \bibnamefont
  {Thew}}, \bibinfo {author} {\bibfnamefont {A.}~\bibnamefont {Ac{\'{i}}n}},
  \bibinfo {author} {\bibfnamefont {H.}~\bibnamefont {Zbinden}}, \ and\
  \bibinfo {author} {\bibfnamefont {N.}~\bibnamefont {Gisin}},\ }\href
  {\doibase 10.1103/PhysRevLett.93.010503} {\bibfield  {journal} {\bibinfo
  {journal} {Physical Review Letters}\ }\textbf {\bibinfo {volume} {93}},\
  \bibinfo {pages} {010503} (\bibinfo {year} {2004})}\BibitemShut {NoStop}%
\bibitem [{\citenamefont {Richart}\ \emph {et~al.}(2012)\citenamefont
  {Richart}, \citenamefont {Fischer},\ and\ \citenamefont
  {Weinfurter}}]{Richart2012}%
  \BibitemOpen
  \bibfield  {author} {\bibinfo {author} {\bibfnamefont {D.}~\bibnamefont
  {Richart}}, \bibinfo {author} {\bibfnamefont {Y.}~\bibnamefont {Fischer}}, \
  and\ \bibinfo {author} {\bibfnamefont {H.}~\bibnamefont {Weinfurter}},\
  }\href {\doibase 10.1007/s00340-011-4854-z} {\bibfield  {journal} {\bibinfo
  {journal} {Applied Physics B}\ }\textbf {\bibinfo {volume} {106}},\ \bibinfo
  {pages} {543} (\bibinfo {year} {2012})}\BibitemShut {NoStop}%
\bibitem [{\citenamefont {Gr\"oblacher}\ \emph {et~al.}(2006)\citenamefont
  {Gr\"oblacher}, \citenamefont {Jennewein}, \citenamefont {Vaziri},
  \citenamefont {Weihs},\ and\ \citenamefont {Zeilinger}}]{Groblacher2006}%
  \BibitemOpen
  \bibfield  {author} {\bibinfo {author} {\bibfnamefont {S.}~\bibnamefont
  {Gr\"oblacher}}, \bibinfo {author} {\bibfnamefont {T.}~\bibnamefont
  {Jennewein}}, \bibinfo {author} {\bibfnamefont {A.}~\bibnamefont {Vaziri}},
  \bibinfo {author} {\bibfnamefont {G.}~\bibnamefont {Weihs}}, \ and\ \bibinfo
  {author} {\bibfnamefont {A.}~\bibnamefont {Zeilinger}},\ }\href {\doibase
  10.1088/1367-2630/8/5/075} {\bibfield  {journal} {\bibinfo  {journal} {New
  Journal of Physics}\ }\textbf {\bibinfo {volume} {8}},\ \bibinfo {pages} {75}
  (\bibinfo {year} {2006})}\BibitemShut {NoStop}%
\bibitem [{\citenamefont {Ali-Khan}\ \emph {et~al.}(2007)\citenamefont
  {Ali-Khan}, \citenamefont {Broadbent},\ and\ \citenamefont
  {Howell}}]{AliKhan2007}%
  \BibitemOpen
  \bibfield  {author} {\bibinfo {author} {\bibfnamefont {I.}~\bibnamefont
  {Ali-Khan}}, \bibinfo {author} {\bibfnamefont {C.~J.}\ \bibnamefont
  {Broadbent}}, \ and\ \bibinfo {author} {\bibfnamefont {J.~C.}\ \bibnamefont
  {Howell}},\ }\href {\doibase 10.1103/PhysRevLett.98.060503} {\bibfield
  {journal} {\bibinfo  {journal} {Phys. Rev. Lett.}\ }\textbf {\bibinfo
  {volume} {98}},\ \bibinfo {pages} {060503} (\bibinfo {year}
  {2007})}\BibitemShut {NoStop}%
\bibitem [{\citenamefont {Mirhosseini}\ \emph {et~al.}(2015)\citenamefont
  {Mirhosseini}, \citenamefont {Magaña-Loaiza}, \citenamefont {O’Sullivan},
  \citenamefont {Rodenburg}, \citenamefont {Malik}, \citenamefont {Lavery},
  \citenamefont {Padgett}, \citenamefont {Gauthier},\ and\ \citenamefont
  {Boyd}}]{Mirhosseini2015}%
  \BibitemOpen
  \bibfield  {author} {\bibinfo {author} {\bibfnamefont {M.}~\bibnamefont
  {Mirhosseini}}, \bibinfo {author} {\bibfnamefont {O.~S.}\ \bibnamefont
  {Magaña-Loaiza}}, \bibinfo {author} {\bibfnamefont {M.~N.}\ \bibnamefont
  {O’Sullivan}}, \bibinfo {author} {\bibfnamefont {B.}~\bibnamefont
  {Rodenburg}}, \bibinfo {author} {\bibfnamefont {M.}~\bibnamefont {Malik}},
  \bibinfo {author} {\bibfnamefont {M.~P.~J.}\ \bibnamefont {Lavery}}, \bibinfo
  {author} {\bibfnamefont {M.~J.}\ \bibnamefont {Padgett}}, \bibinfo {author}
  {\bibfnamefont {D.~J.}\ \bibnamefont {Gauthier}}, \ and\ \bibinfo {author}
  {\bibfnamefont {R.~W.}\ \bibnamefont {Boyd}},\ }\href
  {http://stacks.iop.org/1367-2630/17/i=3/a=033033} {\bibfield  {journal}
  {\bibinfo  {journal} {New Journal of Physics}\ }\textbf {\bibinfo {volume}
  {17}},\ \bibinfo {pages} {033033} (\bibinfo {year} {2015})}\BibitemShut
  {NoStop}%
\bibitem [{\citenamefont {Zhong}\ \emph {et~al.}(2015)\citenamefont {Zhong},
  \citenamefont {Zhou}, \citenamefont {Horansky}, \citenamefont {Lee},
  \citenamefont {Verma}, \citenamefont {Lita}, \citenamefont {Restelli},
  \citenamefont {Bienfang}, \citenamefont {Mirin}, \citenamefont {Gerrits},
  \citenamefont {Nam}, \citenamefont {Marsili}, \citenamefont {Shaw},
  \citenamefont {Zhang}, \citenamefont {Wang}, \citenamefont {Englund},
  \citenamefont {Wornell}, \citenamefont {Shapiro},\ and\ \citenamefont
  {Wong}}]{Zhong2015}%
  \BibitemOpen
  \bibfield  {author} {\bibinfo {author} {\bibfnamefont {T.}~\bibnamefont
  {Zhong}}, \bibinfo {author} {\bibfnamefont {H.}~\bibnamefont {Zhou}},
  \bibinfo {author} {\bibfnamefont {R.~D.}\ \bibnamefont {Horansky}}, \bibinfo
  {author} {\bibfnamefont {C.}~\bibnamefont {Lee}}, \bibinfo {author}
  {\bibfnamefont {V.~B.}\ \bibnamefont {Verma}}, \bibinfo {author}
  {\bibfnamefont {A.~E.}\ \bibnamefont {Lita}}, \bibinfo {author}
  {\bibfnamefont {A.}~\bibnamefont {Restelli}}, \bibinfo {author}
  {\bibfnamefont {J.~C.}\ \bibnamefont {Bienfang}}, \bibinfo {author}
  {\bibfnamefont {R.~P.}\ \bibnamefont {Mirin}}, \bibinfo {author}
  {\bibfnamefont {T.}~\bibnamefont {Gerrits}}, \bibinfo {author} {\bibfnamefont
  {S.~W.}\ \bibnamefont {Nam}}, \bibinfo {author} {\bibfnamefont
  {F.}~\bibnamefont {Marsili}}, \bibinfo {author} {\bibfnamefont {M.~D.}\
  \bibnamefont {Shaw}}, \bibinfo {author} {\bibfnamefont {Z.}~\bibnamefont
  {Zhang}}, \bibinfo {author} {\bibfnamefont {L.}~\bibnamefont {Wang}},
  \bibinfo {author} {\bibfnamefont {D.}~\bibnamefont {Englund}}, \bibinfo
  {author} {\bibfnamefont {G.~W.}\ \bibnamefont {Wornell}}, \bibinfo {author}
  {\bibfnamefont {J.~H.}\ \bibnamefont {Shapiro}}, \ and\ \bibinfo {author}
  {\bibfnamefont {F.~N.~C.}\ \bibnamefont {Wong}},\ }\href {\doibase
  10.1088/1367-2630/17/2/022002} {\bibfield  {journal} {\bibinfo  {journal}
  {New Journal of Physics}\ }\textbf {\bibinfo {volume} {17}},\ \bibinfo
  {pages} {022002} (\bibinfo {year} {2015})}\BibitemShut {NoStop}%
\bibitem [{\citenamefont {Marcikic}\ \emph {et~al.}(2002)\citenamefont
  {Marcikic}, \citenamefont {de~Riedmatten}, \citenamefont {Tittel},
  \citenamefont {Scarani}, \citenamefont {Zbinden},\ and\ \citenamefont
  {Gisin}}]{Marcikic2002}%
  \BibitemOpen
  \bibfield  {author} {\bibinfo {author} {\bibfnamefont {I.}~\bibnamefont
  {Marcikic}}, \bibinfo {author} {\bibfnamefont {H.}~\bibnamefont
  {de~Riedmatten}}, \bibinfo {author} {\bibfnamefont {W.}~\bibnamefont
  {Tittel}}, \bibinfo {author} {\bibfnamefont {V.}~\bibnamefont {Scarani}},
  \bibinfo {author} {\bibfnamefont {H.}~\bibnamefont {Zbinden}}, \ and\
  \bibinfo {author} {\bibfnamefont {N.}~\bibnamefont {Gisin}},\ }\href
  {\doibase 10.1103/PhysRevA.66.062308} {\bibfield  {journal} {\bibinfo
  {journal} {Phys. Rev. A}\ }\textbf {\bibinfo {volume} {66}},\ \bibinfo
  {pages} {62308} (\bibinfo {year} {2002})}\BibitemShut {NoStop}%
\bibitem [{\citenamefont {Sangouard}\ \emph {et~al.}(2011)\citenamefont
  {Sangouard}, \citenamefont {Simon}, \citenamefont {de~Riedmatten},\ and\
  \citenamefont {Gisin}}]{Sangouard2011}%
  \BibitemOpen
  \bibfield  {author} {\bibinfo {author} {\bibfnamefont {N.}~\bibnamefont
  {Sangouard}}, \bibinfo {author} {\bibfnamefont {C.}~\bibnamefont {Simon}},
  \bibinfo {author} {\bibfnamefont {H.}~\bibnamefont {de~Riedmatten}}, \ and\
  \bibinfo {author} {\bibfnamefont {N.}~\bibnamefont {Gisin}},\ }\href
  {\doibase 10.1103/RevModPhys.83.33} {\bibfield  {journal} {\bibinfo
  {journal} {Rev. Mod. Phys.}\ }\textbf {\bibinfo {volume} {83}},\ \bibinfo
  {pages} {33} (\bibinfo {year} {2011})}\BibitemShut {NoStop}%
\bibitem [{\citenamefont {Zhou}\ \emph {et~al.}(2015)\citenamefont {Zhou},
  \citenamefont {Hua}, \citenamefont {Liu}, \citenamefont {Chen}, \citenamefont
  {Xu}, \citenamefont {Han}, \citenamefont {Li},\ and\ \citenamefont
  {Guo}}]{Zhou2015}%
  \BibitemOpen
  \bibfield  {author} {\bibinfo {author} {\bibfnamefont {Z.~Q.}\ \bibnamefont
  {Zhou}}, \bibinfo {author} {\bibfnamefont {Y.~L.}\ \bibnamefont {Hua}},
  \bibinfo {author} {\bibfnamefont {X.}~\bibnamefont {Liu}}, \bibinfo {author}
  {\bibfnamefont {G.}~\bibnamefont {Chen}}, \bibinfo {author} {\bibfnamefont
  {J.~S.}\ \bibnamefont {Xu}}, \bibinfo {author} {\bibfnamefont {Y.~J.}\
  \bibnamefont {Han}}, \bibinfo {author} {\bibfnamefont {C.~F.}\ \bibnamefont
  {Li}}, \ and\ \bibinfo {author} {\bibfnamefont {G.~C.}\ \bibnamefont {Guo}},\
  }\href {\doibase 10.1103/PhysRevLett.115.070502} {\bibfield  {journal}
  {\bibinfo  {journal} {Physical Review Letters}\ }\textbf {\bibinfo {volume}
  {115}},\ \bibinfo {pages} {1} (\bibinfo {year} {2015})}\BibitemShut {NoStop}%
\bibitem [{\citenamefont {Ding}\ \emph {et~al.}(2016)\citenamefont {Ding},
  \citenamefont {Zhang}, \citenamefont {Shi}, \citenamefont {Zhou},
  \citenamefont {Li}, \citenamefont {Shi},\ and\ \citenamefont
  {Guo}}]{Ding2016}%
  \BibitemOpen
  \bibfield  {author} {\bibinfo {author} {\bibfnamefont {D.-S.}\ \bibnamefont
  {Ding}}, \bibinfo {author} {\bibfnamefont {W.}~\bibnamefont {Zhang}},
  \bibinfo {author} {\bibfnamefont {S.}~\bibnamefont {Shi}}, \bibinfo {author}
  {\bibfnamefont {Z.-Y.}\ \bibnamefont {Zhou}}, \bibinfo {author}
  {\bibfnamefont {Y.}~\bibnamefont {Li}}, \bibinfo {author} {\bibfnamefont
  {B.-S.}\ \bibnamefont {Shi}}, \ and\ \bibinfo {author} {\bibfnamefont
  {G.-C.}\ \bibnamefont {Guo}},\ }\href {\doibase 10.1038/lsa.2016.157}
  {\bibfield  {journal} {\bibinfo  {journal} {Light: Science \& Applications}\
  }\textbf {\bibinfo {volume} {5}},\ \bibinfo {pages} {e16157} (\bibinfo {year}
  {2016})}\BibitemShut {NoStop}%
\bibitem [{\citenamefont {Tiranov}\ \emph {et~al.}(2016)\citenamefont
  {Tiranov}, \citenamefont {Strassmann}, \citenamefont {Lavoie}, \citenamefont
  {Brunner}, \citenamefont {Huber}, \citenamefont {Verma}, \citenamefont {Nam},
  \citenamefont {Mirin}, \citenamefont {Lita}, \citenamefont {Marsili},
  \citenamefont {Afzelius}, \citenamefont {Bussi\`eres},\ and\ \citenamefont
  {Gisin}}]{Tiranov2016a}%
  \BibitemOpen
  \bibfield  {author} {\bibinfo {author} {\bibfnamefont {A.}~\bibnamefont
  {Tiranov}}, \bibinfo {author} {\bibfnamefont {P.~C.}\ \bibnamefont
  {Strassmann}}, \bibinfo {author} {\bibfnamefont {J.}~\bibnamefont {Lavoie}},
  \bibinfo {author} {\bibfnamefont {N.}~\bibnamefont {Brunner}}, \bibinfo
  {author} {\bibfnamefont {M.}~\bibnamefont {Huber}}, \bibinfo {author}
  {\bibfnamefont {V.~B.}\ \bibnamefont {Verma}}, \bibinfo {author}
  {\bibfnamefont {S.~W.}\ \bibnamefont {Nam}}, \bibinfo {author} {\bibfnamefont
  {R.~P.}\ \bibnamefont {Mirin}}, \bibinfo {author} {\bibfnamefont {A.~E.}\
  \bibnamefont {Lita}}, \bibinfo {author} {\bibfnamefont {F.}~\bibnamefont
  {Marsili}}, \bibinfo {author} {\bibfnamefont {M.}~\bibnamefont {Afzelius}},
  \bibinfo {author} {\bibfnamefont {F.}~\bibnamefont {Bussi\`eres}}, \ and\
  \bibinfo {author} {\bibfnamefont {N.}~\bibnamefont {Gisin}},\ }\href
  {\doibase 10.1103/PhysRevLett.117.240506} {\bibfield  {journal} {\bibinfo
  {journal} {Phys. Rev. Lett.}\ }\textbf {\bibinfo {volume} {117}},\ \bibinfo
  {pages} {240506} (\bibinfo {year} {2016})}\BibitemShut {NoStop}%
\bibitem [{\citenamefont {Simon}\ \emph {et~al.}(2007)\citenamefont {Simon},
  \citenamefont {de~Riedmatten}, \citenamefont {Afzelius}, \citenamefont
  {Sangouard}, \citenamefont {Zbinden},\ and\ \citenamefont
  {Gisin}}]{Simon2007}%
  \BibitemOpen
  \bibfield  {author} {\bibinfo {author} {\bibfnamefont {C.}~\bibnamefont
  {Simon}}, \bibinfo {author} {\bibfnamefont {H.}~\bibnamefont
  {de~Riedmatten}}, \bibinfo {author} {\bibfnamefont {M.}~\bibnamefont
  {Afzelius}}, \bibinfo {author} {\bibfnamefont {N.}~\bibnamefont {Sangouard}},
  \bibinfo {author} {\bibfnamefont {H.}~\bibnamefont {Zbinden}}, \ and\
  \bibinfo {author} {\bibfnamefont {N.}~\bibnamefont {Gisin}},\ }\href
  {http://link.aps.org/abstract/PRL/v98/e190503} {\bibfield  {journal}
  {\bibinfo  {journal} {Phys. Rev. Lett.}\ }\textbf {\bibinfo {volume} {98}},\
  \bibinfo {pages} {190503} (\bibinfo {year} {2007})}\BibitemShut {NoStop}%
\bibitem [{\citenamefont {Gross}\ \emph {et~al.}(2010)\citenamefont {Gross},
  \citenamefont {Liu}, \citenamefont {Flammia}, \citenamefont {Becker},\ and\
  \citenamefont {Eisert}}]{Gross2010}%
  \BibitemOpen
  \bibfield  {author} {\bibinfo {author} {\bibfnamefont {D.}~\bibnamefont
  {Gross}}, \bibinfo {author} {\bibfnamefont {Y.-K.}\ \bibnamefont {Liu}},
  \bibinfo {author} {\bibfnamefont {S.~T.}\ \bibnamefont {Flammia}}, \bibinfo
  {author} {\bibfnamefont {S.}~\bibnamefont {Becker}}, \ and\ \bibinfo {author}
  {\bibfnamefont {J.}~\bibnamefont {Eisert}},\ }\href {\doibase
  10.1103/PhysRevLett.105.150401} {\bibfield  {journal} {\bibinfo  {journal}
  {Phys. Rev. Lett.}\ }\textbf {\bibinfo {volume} {105}},\ \bibinfo {pages}
  {150401} (\bibinfo {year} {2010})}\BibitemShut {NoStop}%
\bibitem [{\citenamefont {Tonolini}\ \emph {et~al.}(2014)\citenamefont
  {Tonolini}, \citenamefont {Chan}, \citenamefont {Agnew}, \citenamefont
  {Lindsay},\ and\ \citenamefont {Leach}}]{Tonolini2014}%
  \BibitemOpen
  \bibfield  {author} {\bibinfo {author} {\bibfnamefont {F.}~\bibnamefont
  {Tonolini}}, \bibinfo {author} {\bibfnamefont {S.}~\bibnamefont {Chan}},
  \bibinfo {author} {\bibfnamefont {M.}~\bibnamefont {Agnew}}, \bibinfo
  {author} {\bibfnamefont {A.}~\bibnamefont {Lindsay}}, \ and\ \bibinfo
  {author} {\bibfnamefont {J.}~\bibnamefont {Leach}},\ }\href
  {http://dx.doi.org/10.1038/srep06542} {\bibfield  {journal} {\bibinfo
  {journal} {Scientific Reports}\ }\textbf {\bibinfo {volume} {4}},\ \bibinfo
  {pages} {6542 EP } (\bibinfo {year} {2014})}\BibitemShut {NoStop}%
\bibitem [{\citenamefont {Giovannini}\ \emph {et~al.}(2013)\citenamefont
  {Giovannini}, \citenamefont {Romero}, \citenamefont {Leach}, \citenamefont
  {Dudley}, \citenamefont {Forbes},\ and\ \citenamefont
  {Padgett}}]{Giovannini2013}%
  \BibitemOpen
  \bibfield  {author} {\bibinfo {author} {\bibfnamefont {D.}~\bibnamefont
  {Giovannini}}, \bibinfo {author} {\bibfnamefont {J.}~\bibnamefont {Romero}},
  \bibinfo {author} {\bibfnamefont {J.}~\bibnamefont {Leach}}, \bibinfo
  {author} {\bibfnamefont {A.}~\bibnamefont {Dudley}}, \bibinfo {author}
  {\bibfnamefont {A.}~\bibnamefont {Forbes}}, \ and\ \bibinfo {author}
  {\bibfnamefont {M.~J.}\ \bibnamefont {Padgett}},\ }\href {\doibase
  10.1103/PhysRevLett.110.143601} {\bibfield  {journal} {\bibinfo  {journal}
  {Phys. Rev. Lett.}\ }\textbf {\bibinfo {volume} {110}},\ \bibinfo {pages}
  {143601} (\bibinfo {year} {2013})}\BibitemShut {NoStop}%
\bibitem [{\citenamefont {Howland}\ \emph {et~al.}(2016)\citenamefont
  {Howland}, \citenamefont {Knarr}, \citenamefont {Schneeloch}, \citenamefont
  {Lum},\ and\ \citenamefont {Howell}}]{Howland2016}%
  \BibitemOpen
  \bibfield  {author} {\bibinfo {author} {\bibfnamefont {G.~A.}\ \bibnamefont
  {Howland}}, \bibinfo {author} {\bibfnamefont {S.~H.}\ \bibnamefont {Knarr}},
  \bibinfo {author} {\bibfnamefont {J.}~\bibnamefont {Schneeloch}}, \bibinfo
  {author} {\bibfnamefont {D.~J.}\ \bibnamefont {Lum}}, \ and\ \bibinfo
  {author} {\bibfnamefont {J.~C.}\ \bibnamefont {Howell}},\ }\href {\doibase
  10.1103/PhysRevX.6.021018} {\bibfield  {journal} {\bibinfo  {journal} {Phys.
  Rev. X}\ }\textbf {\bibinfo {volume} {6}},\ \bibinfo {pages} {021018}
  (\bibinfo {year} {2016})}\BibitemShut {NoStop}%
\bibitem [{\citenamefont {Erker}\ \emph {et~al.}(2017)\citenamefont {Erker},
  \citenamefont {Krenn},\ and\ \citenamefont {Huber}}]{Erker2017}%
  \BibitemOpen
  \bibfield  {author} {\bibinfo {author} {\bibfnamefont {P.}~\bibnamefont
  {Erker}}, \bibinfo {author} {\bibfnamefont {M.}~\bibnamefont {Krenn}}, \ and\
  \bibinfo {author} {\bibfnamefont {M.}~\bibnamefont {Huber}},\ }\href
  {\doibase 10.22331/q-2017-07-28-22} {\bibfield  {journal} {\bibinfo
  {journal} {{Quantum}}\ }\textbf {\bibinfo {volume} {1}},\ \bibinfo {pages}
  {22} (\bibinfo {year} {2017})}\BibitemShut {NoStop}%
\bibitem [{\citenamefont {Clausen}\ \emph {et~al.}(2014)\citenamefont
  {Clausen}, \citenamefont {Bussi\`{e}res}, \citenamefont {Tiranov},
  \citenamefont {Herrmann}, \citenamefont {Silberhorn}, \citenamefont {Sohler},
  \citenamefont {Afzelius},\ and\ \citenamefont {Gisin}}]{Clausen2014a}%
  \BibitemOpen
  \bibfield  {author} {\bibinfo {author} {\bibfnamefont {C.}~\bibnamefont
  {Clausen}}, \bibinfo {author} {\bibfnamefont {F.}~\bibnamefont
  {Bussi\`{e}res}}, \bibinfo {author} {\bibfnamefont {A.}~\bibnamefont
  {Tiranov}}, \bibinfo {author} {\bibfnamefont {H.}~\bibnamefont {Herrmann}},
  \bibinfo {author} {\bibfnamefont {C.}~\bibnamefont {Silberhorn}}, \bibinfo
  {author} {\bibfnamefont {W.}~\bibnamefont {Sohler}}, \bibinfo {author}
  {\bibfnamefont {M.}~\bibnamefont {Afzelius}}, \ and\ \bibinfo {author}
  {\bibfnamefont {N.}~\bibnamefont {Gisin}},\ }\href {\doibase
  10.1088/1367-2630/16/9/093058} {\bibfield  {journal} {\bibinfo  {journal}
  {New Journal of Physics}\ }\textbf {\bibinfo {volume} {16}},\ \bibinfo
  {pages} {093058} (\bibinfo {year} {2014})}\BibitemShut {NoStop}%
\bibitem [{\citenamefont {Clausen}\ \emph {et~al.}(2010)\citenamefont
  {Clausen}, \citenamefont {Usmani}, \citenamefont {Bussi\`{e}res},
  \citenamefont {Sangouard}, \citenamefont {Afzelius}, \citenamefont
  {de~Riedmatten},\ and\ \citenamefont {Gisin}}]{Clausen2011}%
  \BibitemOpen
  \bibfield  {author} {\bibinfo {author} {\bibfnamefont {C.}~\bibnamefont
  {Clausen}}, \bibinfo {author} {\bibfnamefont {I.}~\bibnamefont {Usmani}},
  \bibinfo {author} {\bibfnamefont {F.}~\bibnamefont {Bussi\`{e}res}}, \bibinfo
  {author} {\bibfnamefont {N.}~\bibnamefont {Sangouard}}, \bibinfo {author}
  {\bibfnamefont {M.}~\bibnamefont {Afzelius}}, \bibinfo {author}
  {\bibfnamefont {H.}~\bibnamefont {de~Riedmatten}}, \ and\ \bibinfo {author}
  {\bibfnamefont {N.}~\bibnamefont {Gisin}},\ }\href {\doibase
  10.1038/nature09662} {\bibfield  {journal} {\bibinfo  {journal} {Nature}\
  }\textbf {\bibinfo {volume} {469}},\ \bibinfo {pages} {508} (\bibinfo {year}
  {2010})}\BibitemShut {NoStop}%
\bibitem [{\citenamefont {Franson}(1989)}]{Franson1989}%
  \BibitemOpen
  \bibfield  {author} {\bibinfo {author} {\bibfnamefont {J.~D.}\ \bibnamefont
  {Franson}},\ }\href {\doibase 10.1103/PhysRevLett.62.2205} {\bibfield
  {journal} {\bibinfo  {journal} {Phys. Rev. Lett.}\ }\textbf {\bibinfo
  {volume} {62}},\ \bibinfo {pages} {2205} (\bibinfo {year}
  {1989})}\BibitemShut {NoStop}%
\bibitem [{\citenamefont {Tiranov}\ \emph {et~al.}(2015)\citenamefont
  {Tiranov}, \citenamefont {Lavoie}, \citenamefont {Ferrier}, \citenamefont
  {Goldner}, \citenamefont {Verma}, \citenamefont {Nam}, \citenamefont {Mirin},
  \citenamefont {Lita}, \citenamefont {Marsili}, \citenamefont {Herrmann},
  \citenamefont {Silberhorn}, \citenamefont {Gisin}, \citenamefont {Afzelius},\
  and\ \citenamefont {Bussi\`{e}res}}]{Tiranov2015a}%
  \BibitemOpen
  \bibfield  {author} {\bibinfo {author} {\bibfnamefont {A.}~\bibnamefont
  {Tiranov}}, \bibinfo {author} {\bibfnamefont {J.}~\bibnamefont {Lavoie}},
  \bibinfo {author} {\bibfnamefont {A.}~\bibnamefont {Ferrier}}, \bibinfo
  {author} {\bibfnamefont {P.}~\bibnamefont {Goldner}}, \bibinfo {author}
  {\bibfnamefont {V.~B.}\ \bibnamefont {Verma}}, \bibinfo {author}
  {\bibfnamefont {S.~W.}\ \bibnamefont {Nam}}, \bibinfo {author} {\bibfnamefont
  {R.~P.}\ \bibnamefont {Mirin}}, \bibinfo {author} {\bibfnamefont {A.~E.}\
  \bibnamefont {Lita}}, \bibinfo {author} {\bibfnamefont {F.}~\bibnamefont
  {Marsili}}, \bibinfo {author} {\bibfnamefont {H.}~\bibnamefont {Herrmann}},
  \bibinfo {author} {\bibfnamefont {C.}~\bibnamefont {Silberhorn}}, \bibinfo
  {author} {\bibfnamefont {N.}~\bibnamefont {Gisin}}, \bibinfo {author}
  {\bibfnamefont {M.}~\bibnamefont {Afzelius}}, \ and\ \bibinfo {author}
  {\bibfnamefont {F.}~\bibnamefont {Bussi\`{e}res}},\ }\href {\doibase
  10.1364/OPTICA.2.000279} {\bibfield  {journal} {\bibinfo  {journal} {Optica}\
  }\textbf {\bibinfo {volume} {2}},\ \bibinfo {pages} {279} (\bibinfo {year}
  {2015})}\BibitemShut {NoStop}%
\bibitem [{\citenamefont {Huber}\ and\ \citenamefont
  {de~Vicente}(2013)}]{Huber2013}%
  \BibitemOpen
  \bibfield  {author} {\bibinfo {author} {\bibfnamefont {M.}~\bibnamefont
  {Huber}}\ and\ \bibinfo {author} {\bibfnamefont {J.~I.}\ \bibnamefont
  {de~Vicente}},\ }\href {\doibase 10.1103/PhysRevLett.110.030501} {\bibfield
  {journal} {\bibinfo  {journal} {Physical Review Letters}\ }\textbf {\bibinfo
  {volume} {110}},\ \bibinfo {pages} {030501} (\bibinfo {year}
  {2013})}\BibitemShut {NoStop}%
\bibitem [{\citenamefont {Wootters}(2001)}]{Wootters2001}%
  \BibitemOpen
  \bibfield  {author} {\bibinfo {author} {\bibfnamefont {W.~K.}\ \bibnamefont
  {Wootters}},\ }\href {http://dl.acm.org/citation.cfm?id=2011326.2011329}
  {\bibfield  {journal} {\bibinfo  {journal} {Quantum Info. Comput.}\ }\textbf
  {\bibinfo {volume} {1}},\ \bibinfo {pages} {27} (\bibinfo {year}
  {2001})}\BibitemShut {NoStop}%
\bibitem [{\citenamefont {Martin}\ \emph {et~al.}(2017)\citenamefont {Martin},
  \citenamefont {Guerreiro}, \citenamefont {Tiranov}, \citenamefont
  {Designolle}, \citenamefont {Fr\"owis}, \citenamefont {Brunner},
  \citenamefont {Huber},\ and\ \citenamefont {Gisin}}]{Martin2017}%
  \BibitemOpen
  \bibfield  {author} {\bibinfo {author} {\bibfnamefont {A.}~\bibnamefont
  {Martin}}, \bibinfo {author} {\bibfnamefont {T.}~\bibnamefont {Guerreiro}},
  \bibinfo {author} {\bibfnamefont {A.}~\bibnamefont {Tiranov}}, \bibinfo
  {author} {\bibfnamefont {S.}~\bibnamefont {Designolle}}, \bibinfo {author}
  {\bibfnamefont {F.}~\bibnamefont {Fr\"owis}}, \bibinfo {author}
  {\bibfnamefont {N.}~\bibnamefont {Brunner}}, \bibinfo {author} {\bibfnamefont
  {M.}~\bibnamefont {Huber}}, \ and\ \bibinfo {author} {\bibfnamefont
  {N.}~\bibnamefont {Gisin}},\ }\href {\doibase 10.1103/PhysRevLett.118.110501}
  {\bibfield  {journal} {\bibinfo  {journal} {Phys. Rev. Lett.}\ }\textbf
  {\bibinfo {volume} {118}},\ \bibinfo {pages} {110501} (\bibinfo {year}
  {2017})}\BibitemShut {NoStop}%
\bibitem [{\citenamefont {Jobez}\ \emph {et~al.}(2016)\citenamefont {Jobez},
  \citenamefont {Timoney}, \citenamefont {Laplane}, \citenamefont {Etesse},
  \citenamefont {Ferrier}, \citenamefont {Goldner}, \citenamefont {Gisin},\
  and\ \citenamefont {Afzelius}}]{Jobez2016}%
  \BibitemOpen
  \bibfield  {author} {\bibinfo {author} {\bibfnamefont {P.}~\bibnamefont
  {Jobez}}, \bibinfo {author} {\bibfnamefont {N.}~\bibnamefont {Timoney}},
  \bibinfo {author} {\bibfnamefont {C.}~\bibnamefont {Laplane}}, \bibinfo
  {author} {\bibfnamefont {J.}~\bibnamefont {Etesse}}, \bibinfo {author}
  {\bibfnamefont {A.}~\bibnamefont {Ferrier}}, \bibinfo {author} {\bibfnamefont
  {P.}~\bibnamefont {Goldner}}, \bibinfo {author} {\bibfnamefont
  {N.}~\bibnamefont {Gisin}}, \ and\ \bibinfo {author} {\bibfnamefont
  {M.}~\bibnamefont {Afzelius}},\ }\href {\doibase 10.1103/PhysRevA.93.032327}
  {\bibfield  {journal} {\bibinfo  {journal} {Phys. Rev. A}\ }\textbf {\bibinfo
  {volume} {93}},\ \bibinfo {pages} {032327} (\bibinfo {year}
  {2016})}\BibitemShut {NoStop}%
\bibitem [{\citenamefont {Sinclair}\ \emph {et~al.}(2014)\citenamefont
  {Sinclair}, \citenamefont {Saglamyurek}, \citenamefont {Mallahzadeh},
  \citenamefont {Slater}, \citenamefont {George}, \citenamefont {Ricken},
  \citenamefont {Hedges}, \citenamefont {Oblak}, \citenamefont {Simon},
  \citenamefont {Sohler},\ and\ \citenamefont {Tittel}}]{Sinclair2014}%
  \BibitemOpen
  \bibfield  {author} {\bibinfo {author} {\bibfnamefont {N.}~\bibnamefont
  {Sinclair}}, \bibinfo {author} {\bibfnamefont {E.}~\bibnamefont
  {Saglamyurek}}, \bibinfo {author} {\bibfnamefont {H.}~\bibnamefont
  {Mallahzadeh}}, \bibinfo {author} {\bibfnamefont {J.~A.}\ \bibnamefont
  {Slater}}, \bibinfo {author} {\bibfnamefont {M.}~\bibnamefont {George}},
  \bibinfo {author} {\bibfnamefont {R.}~\bibnamefont {Ricken}}, \bibinfo
  {author} {\bibfnamefont {M.~P.}\ \bibnamefont {Hedges}}, \bibinfo {author}
  {\bibfnamefont {D.}~\bibnamefont {Oblak}}, \bibinfo {author} {\bibfnamefont
  {C.}~\bibnamefont {Simon}}, \bibinfo {author} {\bibfnamefont
  {W.}~\bibnamefont {Sohler}}, \ and\ \bibinfo {author} {\bibfnamefont
  {W.}~\bibnamefont {Tittel}},\ }\href {\doibase
  10.1103/PhysRevLett.113.053603} {\bibfield  {journal} {\bibinfo  {journal}
  {Phys. Rev. Lett.}\ }\textbf {\bibinfo {volume} {113}},\ \bibinfo {pages}
  {053603} (\bibinfo {year} {2014})}\BibitemShut {NoStop}%
\bibitem [{\citenamefont {Moroder}\ \emph {et~al.}(2013)\citenamefont
  {Moroder}, \citenamefont {Bancal}, \citenamefont {Liang}, \citenamefont
  {Hofmann},\ and\ \citenamefont {G\"uhne}}]{Moroder2013}%
  \BibitemOpen
  \bibfield  {author} {\bibinfo {author} {\bibfnamefont {T.}~\bibnamefont
  {Moroder}}, \bibinfo {author} {\bibfnamefont {J.-D.}\ \bibnamefont {Bancal}},
  \bibinfo {author} {\bibfnamefont {Y.-C.}\ \bibnamefont {Liang}}, \bibinfo
  {author} {\bibfnamefont {M.}~\bibnamefont {Hofmann}}, \ and\ \bibinfo
  {author} {\bibfnamefont {O.}~\bibnamefont {G\"uhne}},\ }\href {\doibase
  10.1103/PhysRevLett.111.030501} {\bibfield  {journal} {\bibinfo  {journal}
  {Phys. Rev. Lett.}\ }\textbf {\bibinfo {volume} {111}},\ \bibinfo {pages}
  {030501} (\bibinfo {year} {2013})}\BibitemShut {NoStop}%
\bibitem [{\citenamefont {Brunner}\ \emph {et~al.}(2008)\citenamefont
  {Brunner}, \citenamefont {Pironio}, \citenamefont {Acin}, \citenamefont
  {Gisin}, \citenamefont {M\'ethot},\ and\ \citenamefont
  {Scarani}}]{Brunner2008}%
  \BibitemOpen
  \bibfield  {author} {\bibinfo {author} {\bibfnamefont {N.}~\bibnamefont
  {Brunner}}, \bibinfo {author} {\bibfnamefont {S.}~\bibnamefont {Pironio}},
  \bibinfo {author} {\bibfnamefont {A.}~\bibnamefont {Acin}}, \bibinfo {author}
  {\bibfnamefont {N.}~\bibnamefont {Gisin}}, \bibinfo {author} {\bibfnamefont
  {A.~A.}\ \bibnamefont {M\'ethot}}, \ and\ \bibinfo {author} {\bibfnamefont
  {V.}~\bibnamefont {Scarani}},\ }\href {\doibase
  10.1103/PhysRevLett.100.210503} {\bibfield  {journal} {\bibinfo  {journal}
  {Phys. Rev. Lett.}\ }\textbf {\bibinfo {volume} {100}},\ \bibinfo {pages}
  {210503} (\bibinfo {year} {2008})}\BibitemShut {NoStop}%
\bibitem [{\citenamefont {Min\'a\v{r}}\ \emph {et~al.}(2008)\citenamefont
  {Min\'a\v{r}}, \citenamefont {de~Riedmatten}, \citenamefont {Simon},
  \citenamefont {Zbinden},\ and\ \citenamefont {Gisin}}]{Minar2008}%
  \BibitemOpen
  \bibfield  {author} {\bibinfo {author} {\bibfnamefont {J.}~\bibnamefont
  {Min\'a\v{r}}}, \bibinfo {author} {\bibfnamefont {H.}~\bibnamefont
  {de~Riedmatten}}, \bibinfo {author} {\bibfnamefont {C.}~\bibnamefont
  {Simon}}, \bibinfo {author} {\bibfnamefont {H.}~\bibnamefont {Zbinden}}, \
  and\ \bibinfo {author} {\bibfnamefont {N.}~\bibnamefont {Gisin}},\ }\href
  {\doibase 10.1103/PhysRevA.77.052325} {\bibfield  {journal} {\bibinfo
  {journal} {Phys. Rev. A}\ }\textbf {\bibinfo {volume} {77}},\ \bibinfo
  {pages} {052325} (\bibinfo {year} {2008})}\BibitemShut {NoStop}%
\end{thebibliography}%

\newpage

\clearpage

\appendix

\section*{Appendix for ``\TitleName''}

In this \suppmat{} we provide more details about experimental results and theoretical method that was implemented to quantify multi-dimensional entanglement.

\section{Details of experimental results}

We have performed 12 complete experiments following the method explained in the main text. Here we describe in more details one of these runs and provide details about final results.

\begin{figure*}[t!]
\centering
\includegraphics[width=0.8\textwidth]{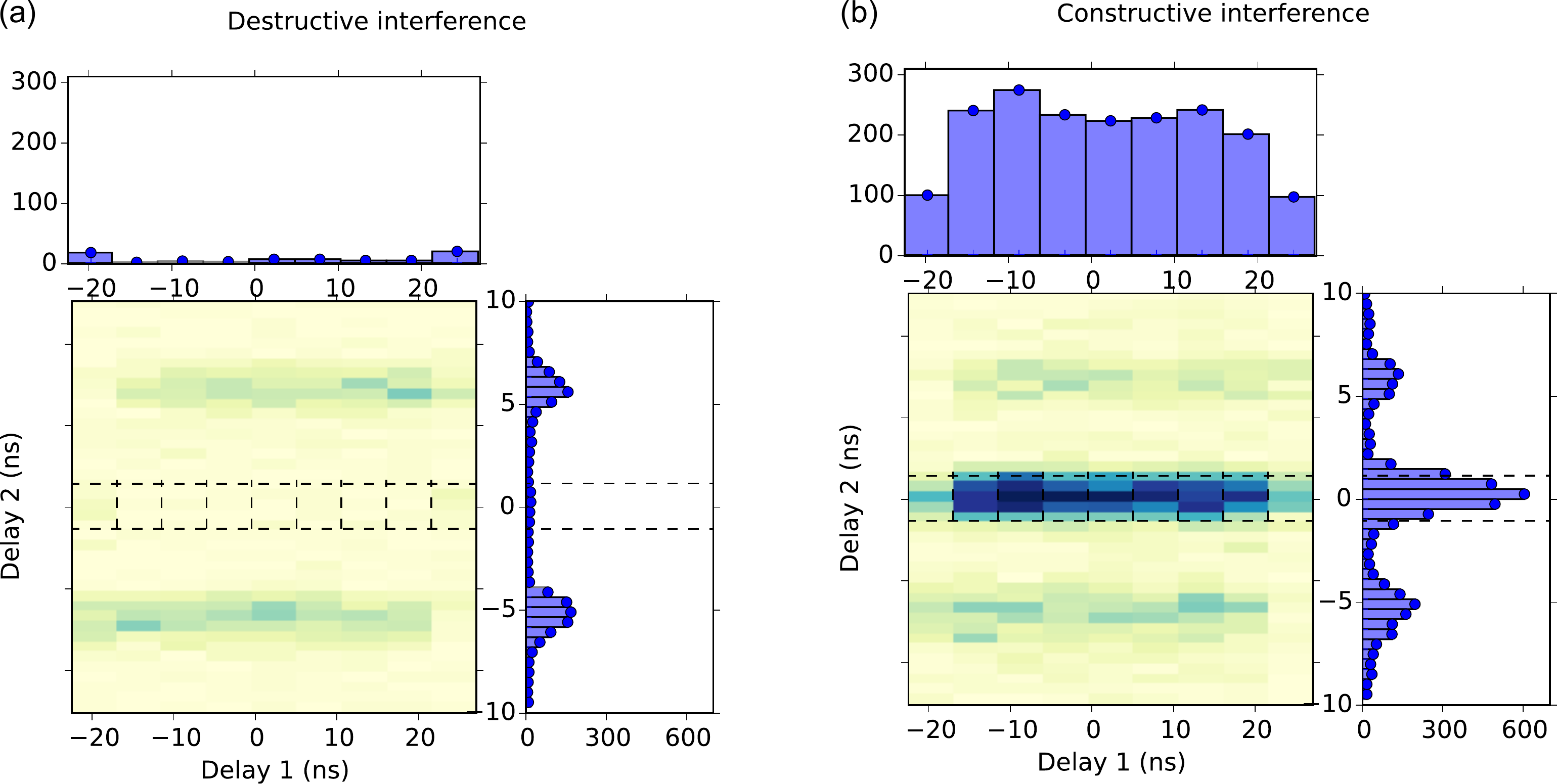}
\caption{Example of the interference measurement corresponding to destructive and constructive interference between neighboring temporal modes. From this measurement we extract the visibilities corresponding to different pairs of neighboring temporal modes.
}
\label{figsm:2dvis}
\end{figure*}

First, we provide details about the measurement of the visibility between the neighboring temporal modes. Fig.~\ref{figsm:2dvis} illustrates 2D image representing the coincidence measurement for different temporal modes (Delay~1) as a function of delay between two detectors $D_s$ and $D_i$ (Delay~2). 
The coincidence histograms between detectors $D_s$ and $D_i$ shows three peaks corresponding to different path combinations for travelling idler and signal photon after storage. By varying the phase of the interferometer $\phi_s$ we observe the interference for central peak which represents post-selected time-bin entangled state
\begin{align}
\ket{\Phi_d} = \frac{1}{\sqrt{d}}\sum_{j=1}^d c_j\ket{j}_i \otimes \ket{j}_s\ .
\label{eqsm:qudit}
\end{align}
The separation between peaks is equal to the travel-time difference between different arms of the interferometer $\Delta=$5.5~ns (Fig.~\ref{figsm:2dvis}). The central peak is post-selected using 3~ns temporal window (Delay~2) illustrated by dashed line. We define different temporal modes by discretizing temporal pulse using period $\Delta$ (Delay~1 in Fig.~\ref{figsm:2dvis}).

We measure the visibility for each pair of neighboring temporal modes by comparing number of coincidences corresponding to destructive (Fig.~\ref{figsm:2dvis}(a)) and constructive (Fig.~\ref{figsm:2dvis}(b)) interferences between different temporal modes. The visibility is reduced at the edges of the pulse which can be seen from increased number of coincidences for destructive interference for first and last histogram bin (Fig.~\ref{figsm:2dvis}). This is explained by fast intensity variation at the beginning and at the end of the pulse which reduces the maximum achievable visibility. To measure intensity of each temporal mode $c_j$ we block one of the arms of the signal interferometer and repeat coincidence measurement described above.

\begin{figure}[t!]
\centering
\includegraphics[width=0.4\textwidth]{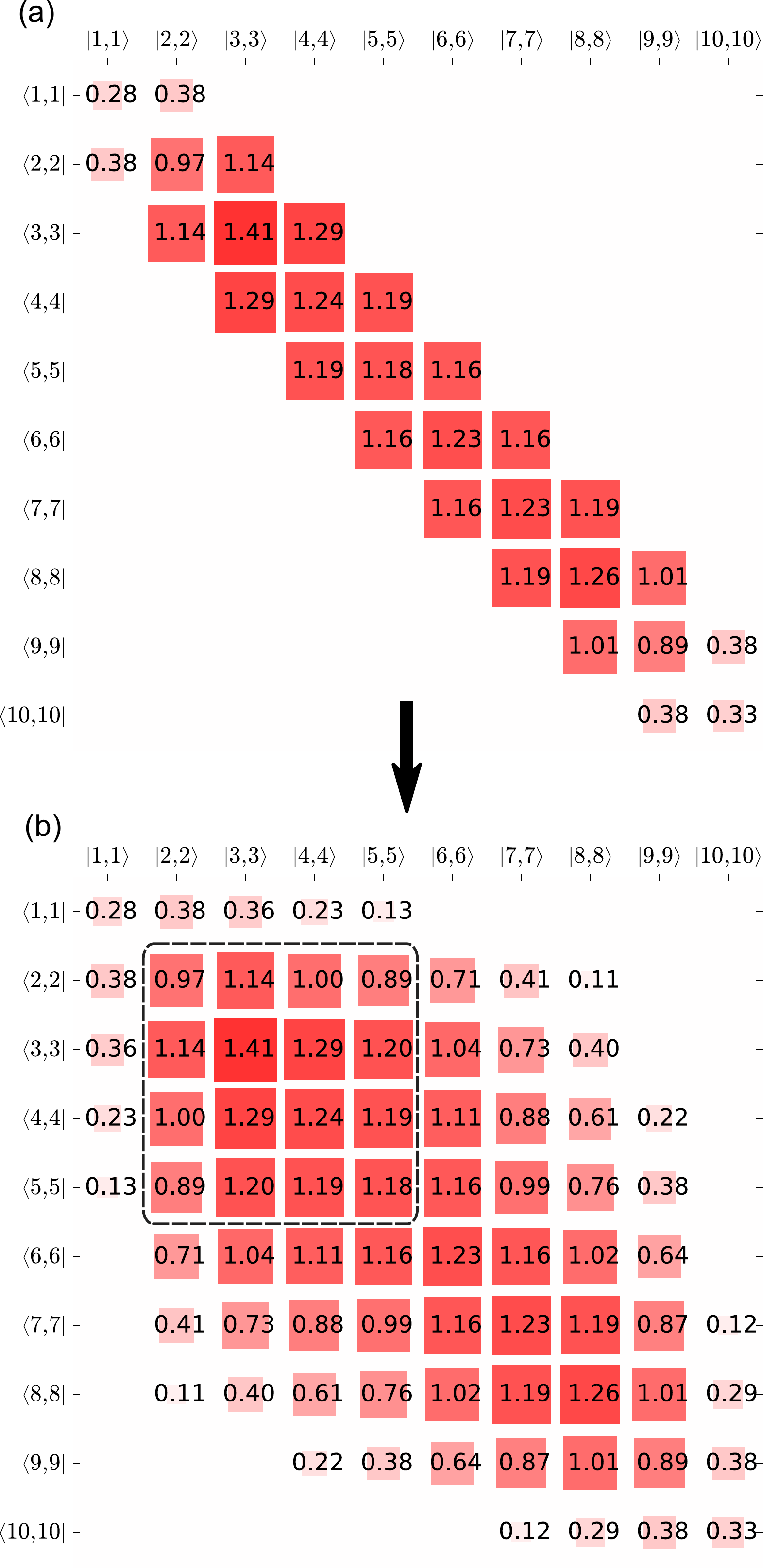}
\caption{Illustration of the sub matrix reconstruction for experimental data of one run. The values were normalized with respect to the maximum dimension~(10 in this case).
}
\label{figsm:dms}
\end{figure}

After we show the full sub-matrix reconstruction. Fig.~\ref{figsm:dms} shows the sub-matrix before and after application of the theoretical method. Only elements from the first $r_{j,j}$ and second $r_{j,j+1}$ diagonals were measured experimentally using a pair of interferometers. Application of the method based on the positivity of the density matrix (described in the main text) gives a lower bound on the elements for all other diagonals ($r_{j,j+2}$, $r_{j,j+3}$ and so on). These elements are further used to give a lower bound on the entanglement of formation $E_{oF}$ based on expression from Ref.~\cite{Huber2013}
\begin{equation}
E_{oF} \geq - \log_2(1-\frac{B^2}{2})\ ,
\label{eqn:eof}
\end{equation}
where we have defined the quantity $B$ as 
\begin{equation}
\frac{2}{\sqrt{|C|}} \left(\sum_{(j,k)\in C \atop j < k}|\bra{j,j}\rho\ket{k,k}|-\sqrt{\bra{j,k}\rho\ket{j,k}\bra{k,j}\rho\ket{k,j}} \right ).
\label{eqn:b}
\end{equation}
The sub-matrix which gives maximum value of $E_{oF}$ is indicated by a dashed line.

\begin{figure}
\centering
\includegraphics[width=0.4\textwidth]{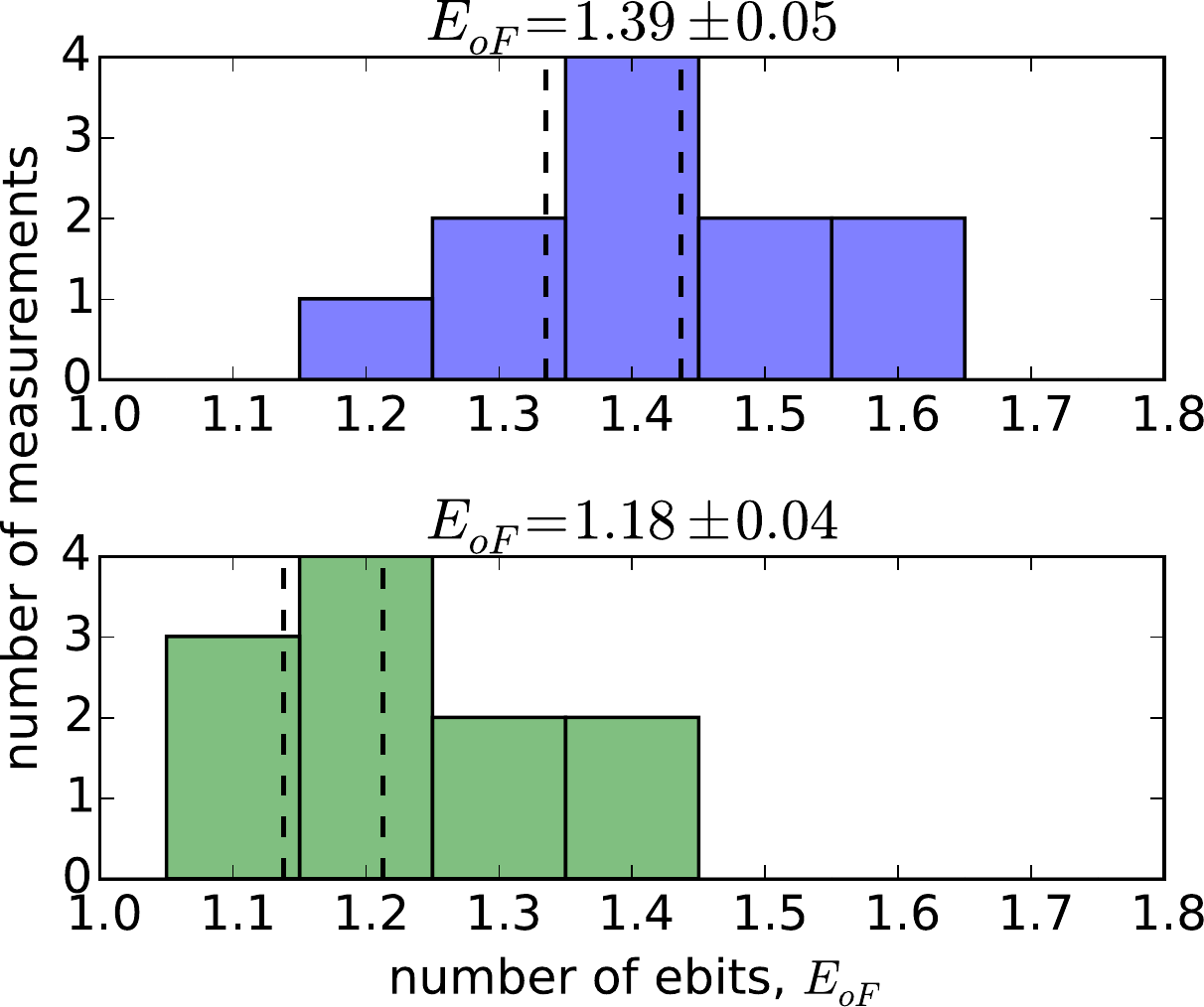}
\caption{Summary of all results without (a) and including white noise contribution (b).
}
\label{figsm:sum}
\end{figure}

Same set of data was accumulated many times and for each the certified number of \ebits{} was calculated. Due to the instability of each interferometer the visibility value varied from time to time. The average visibility of 97\% was measured. Final certified values of entanglement of formation are given in Fig.~\ref{figsm:sum} and are all above one for both cases without or including white noise contribution which modifies the diagonal element of the density matrix and reduces certified number of \ebits{}.

\section{Bounds on entanglement: noise sensitivity}

\begin{figure}[ht!]
\centering
\includegraphics[width=0.4\textwidth]{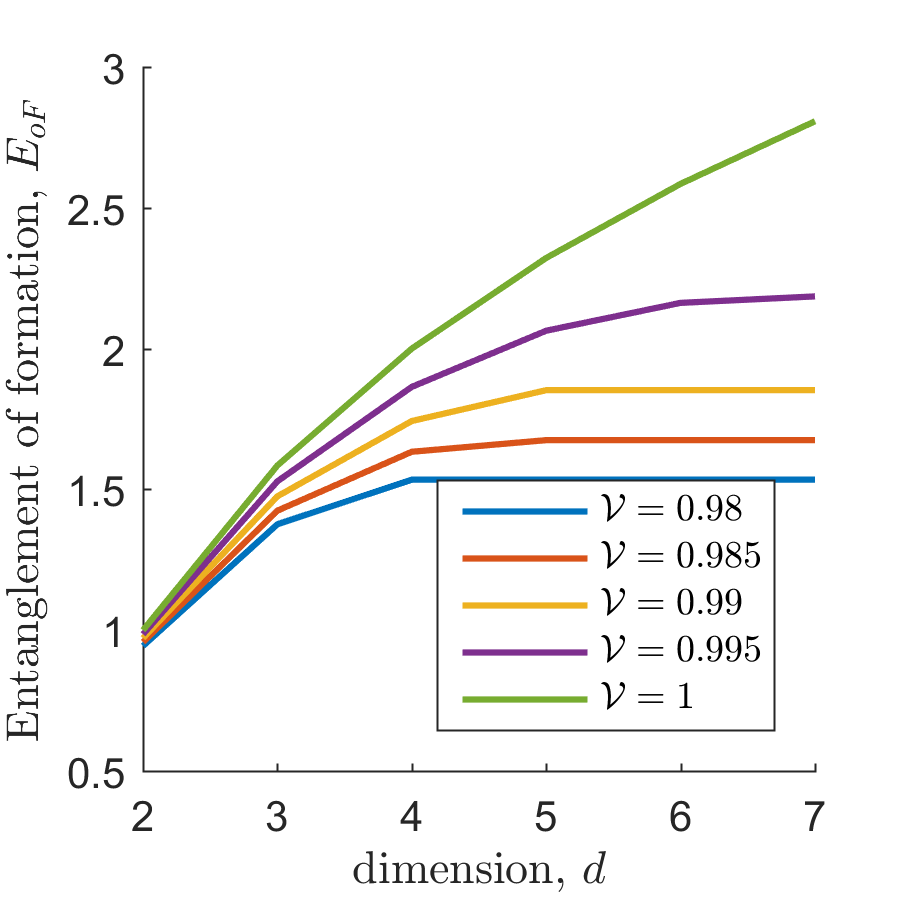}
\caption{Lower bound on the entanglement of formation (in terms of number of ebits) as a function of the dimension $d$, for visibilities $ \mathcal{V}$ from 1 to 0.98. }
\label{figsm:ebits}
\end{figure}

We discuss here in more details the characterization of the density matrix obtained via the method described in the main text. In particular we consider again the situation expected from our experiment, that is, $r_{j,j}=1$ and $r_{j,j+1}=\mathcal{V}$. Applying the method, we get the following bounds for the (unmeasured) coherence terms
\begin{equation}
\begin{array}{rcl}
r_{j,j+2} & \geq & 2\mathcal{V}^2-1 \\ 
r_{j,j+3} & \geq & \mathcal{V}(4\mathcal{V}^2-3) \\ 
r_{j,j+4} & \geq & 8\mathcal{V}^4-8\mathcal{V}^2+1 \\ 
r_{j,j+5} & \geq & \mathcal{V}(16\mathcal{V}^4-20\mathcal{V}^2+5).
\end{array}
\end{equation}
Based on these bounds, one can then obtain a lower on the entanglement of formation, as discussed in the main text. In Fig. (\ref{figsm:ebits}) we plot the obtained bound on $E_{oF}$ as a function of the dimension of the quantum state. We consider different values of the visibility $\mathcal{V}$. For $\mathcal{V}=1$, one obtains $E_{oF} = \log_2(d)$, which corresponds to the maximally entangled state of dimension $d \times d$, i.e. $\ket{\Phi_d}$. Notice that the maximally entangled state is here the only quantum state compatible with the requirement that $\mathcal{V}=1$. For $\mathcal{V}<1$, one can see that the bound on $E_{oF}$ reaches a maximum (for some dimension, which depends on the value of $\mathcal{V}$) and then remains constant. Hence for limited visibility $\mathcal{V}<1$, there is a maximal amount of $E_{oF} $ that can be certified, independent of the Hilbert space dimension.

\section{Physical model of the phase noise}

The visibility measured for bigger interferometric delays will monotonically decrease due to the finite linewidth of the pump laser.
The phase noise of the pump laser can be approximated by gaussian distribution with standard deviation $\delta\phi$. In this case the visibility scales as $\mathcal{V} \sim e^{-\delta\phi^2/2}$ \citep{Minar2008}. In our case for different temporal modes separated by $n\Delta$ delay we can rewrite it as
\begin{equation}
\mathcal{V}_n = \mathcal{V}_1e^{-2(\pi\delta\nu n \Delta)^2},
\end{equation}
where $\delta\nu$ is the spectral linewidth of the pump laser and $\mathcal{V}_1$ is the visibility between neighboring modes. 
 
Assuming a full-width half-maximum linewidth of the pump laser of 1~kHz and a maximum delay between temporal modes of 50~ns, the expected visibility remains almost constant decreasing only by a factor~0.999. This verifies our approximation of coherent sum between all temporal modes generated and stored in the quantum memory.



%
%
%



\end{document}